\documentclass[11pt,reqno]{article}
\usepackage{amsmath}
\usepackage{amsfonts}
\usepackage{amssymb}
\usepackage{latexsym}
\usepackage[dvips]{graphicx}
\usepackage{epsf}
\usepackage{color}

\usepackage{url}

\allowdisplaybreaks \numberwithin{equation}{section}

\newcommand{\be}{\begin{equation}}
\newcommand{\ee}{\end{equation}}
\newcommand{\bi}{\begin{itemize}}
\newcommand{\ei}{\end{itemize}}
\newcommand{\bea}{\begin{eqnarray}}
\newcommand{\eea}{\end{eqnarray}}
\newcommand{\nn}{\nonumber}

\let\a=\alpha \let\b=\beta  \let\g=\gamma  \let\d=\delta
        \let\l=\lambda
\let\m=\mu    \let\n=\nu          \let\r=\rho
\let\s=\sigma \let\t=\tau     
    
\let\G=\Gamma \let\D=\Delta    
\let\P=\Pi         
\let\O=\Omega  \let\eps=\varepsilon

\let\==\equiv

\let\pa=\partial

\newcommand{\tr}{{\rm tr}}
\newcommand{\Tr}{{\rm Tr}}

\newcommand{\unit}{{\bf{1}}}
\newcommand{\half}{\tfrac{1}{2}}

\newcommand{\cC}{\mathcal{C}}

\newcommand{\cE}{\mathcal{E}}

\newcommand{\cO}{\mathcal{O}}
\newcommand{\cR}{\mathcal{R}}

\newcommand{\e}{\mathrm{e}}

\newcommand{\p}{\partial}

\begin{document}

\thispagestyle{empty}
\begin{flushright} \small
MZ-TH/11-43
\end{flushright}
\bigskip

\begin{center}
{\LARGE\bfseries   Off-diagonal heat-kernel expansion and \\[1.5ex]
\LARGE\bfseries   its application to fields with differential constraints  
}
\\[10mm]
Kai Groh, Frank Saueressig and Omar Zanusso \\[3mm]
{\small\slshape
Institute of Physics, University of Mainz\\
Staudingerweg 7, D-55099 Mainz, Germany \\[1.1ex]
{\upshape\ttfamily kgroh@thep.physik.uni-mainz.de} \\
{\upshape\ttfamily saueressig@thep.physik.uni-mainz.de} \\
{\upshape\ttfamily zanusso@thep.physik.uni-mainz.de} }\\
\end{center}
\vspace{10mm}

\hrule\bigskip

\centerline{\bfseries Abstract} \medskip
\noindent
The off-diagonal heat-kernel expansion of a Laplace operator including a general gauge-connection is computed on a compact manifold without boundary up to third order in the curvatures. These results are used to study the early-time expansion of the traced heat-kernel on the space of transverse vector fields satisfying the differential constraint $D^\mu v_\mu = 0$. It is shown that the resulting 
Seeley-deWitt coefficients generically develop singularities, which vanish if the metric is flat or satisfies the Einstein condition. The implications of our findings for the evaluation of the gravitational functional renormalization group equation are briefly discussed.
\bigskip
\hrule\bigskip
\newpage

\section{Introduction}

The heat-kernel technique is a mathematical tool with a wide range of applicability both in mathematics and theoretical physics \cite{Vassilevich:2003xt, Barvinsky:1985an, Avramidi:2000bm}. This is owed to the fact that it provides a formal way to treat functional traces and determinants of local differential operators, including any Laplace-type (second order differential) operator. Irrespective of the particular context they appear in, the arguments of such traces can often be written in an exponentiated form, that allows their systematic treatment in terms of the heat-kernel
\be\label{hkdef}
H(s) = \e^{-s\D_0} \, .
\ee
Here $\D_0$ denotes a generalized Laplace operator on a closed and torsionless Riemannian manifold that we will refer to as ``spacetime''. In the context of theoretical physics, it allows to give, for example, a definition of the propagator\footnote{Using the well known Schwinger-trick $\frac{1}{\D} = \int_0^\infty ds\; \e^{-s\D}$.} and 1-loop effective actions.\footnote{$\G_{\rm 1-loop} = \half\Tr\log \D = \half\int_0^\infty \frac{ds}{s} \Tr [\e^{-s\D}]$.} Moreover, it allows to compute counterterms and anomalies in an elegant way that can be extended naturally to field theories on arbitrary curved space.
Furthermore, it constitutes an essential ingredient in solving the gravitational functional renormalization group equation \cite{Reuter:1996cp}, see  \cite{Reuter:2007rv,Codello:2008vh} for reviews.

The trace of the heat-kernel has two main expansion schemes.
One is the local or early-time expansion in terms of powers of $s$
\be\label{sdw}
\Tr\;H(s) = \frac{1}{\left(4\pi s\right)^{d/2}} \sum_{n\ge 0} \int d^d x \sqrt{g} s^n \,{\rm tr}\, \overline{A_n} \, ,
\ee
generally referred to as Seeley-deWitt expansion. The scalar quantities $\overline{A_n}(x)$ are called heat-kernel coefficients and are local functions of the curvature invariants and their covariant derivatives.
The other scheme is based on a non-local expansion in terms of curvature tensors and schematically represented by \cite{Barvinsky:1987uw}
\be\label{bv}
\Tr\;H(s) = \frac{1}{\left(4\pi s\right)^{d/2}} \sum_n s^n \int \prod_{i=1}^n \Big( d^d x_i\sqrt{g(x_i)} \Big) \,
F\left(s\D_{i_1},\dots,s\D_{i_n}\right) \cR_{i_1} \dots \cR_{i_n} \, .
\ee
This expansion involves arbitrary powers of derivatives at every order in the curvature tensors $\cR_i=\cR_i(x_i)$.
In (\ref{bv}) each operator $\D_i$ is acting only on the corresponding invariant $\cR_i$ and the tensor structure of the invariants has been suppressed for brevity.
Throughout this paper, we will work with the early-time expansion \eqref{sdw}, since its coefficients and their derivatives can be computed recursively.

In quantum field theory (QFT), the computation of \eqref{hkdef} plays a central role, since it allows explicitly covariant computations of which the aforementioned propagator and $1$-loop effective action are only special examples \cite{DeWitt:1965jb}. In fact, a general trace involving a function of $\D_0$ can formally be related to the
traced heat-kernel by a Laplace transform
\be\label{arbitrary}
 \Tr\; f\!\left(\Delta_0\right) = \int_0^\infty ds\, \tilde{f}\left(s\right)\, \Tr\; H(s)\, .
\ee
Thus, a wide class of covariant computations in QFT, involving different functions $f(x)$ of the same differential operator $\Delta_0$, are reduced to the calculation of the single object $\Tr\; H(s)$.

The caveat of this approach is that the typical operators $\D_0$ of physical interest are often very complicated and, in particular, not of Laplace-type. A prototypical example
is Yang-Mills theory where the  gauge fixed inverse propagator of a gauge field $A_\m^c$ is given by a differential operator that has indices on the internal gauge group as well as spacetime,
and, for a general gauge-parameter $\alpha$, is of non-minimal form in the sense that it includes uncontracted derivatives
\be\label{nonminimal_operator}
\D^{ac}_{\m\n} = - g_{\m\n}D_\a^{ab}D^{\a bc} +\left( 1-\alpha^{-1} \right) D_\m^{ab}D_\n^{bc} \; , \qquad {\rm with} \qquad D_\m^{ab} = \pa_\m \d^{ab} + f^{abc} A_\m^c \, .
\ee
One possibility to simplify such types of operators is the use of covariant projectors \cite{Reuter:1993kw,Gies:2002af,Benedetti:2010nr}
which leads to Laplace-type differential operators acting on subspaces of the original field space.
A typical example for such projections are the transverse decomposition of a vector field, or the transverse traceless decomposition of a symmetric tensor field \cite{York:1973ia}.
Alternatively, one may
carry out a resummation of the non-minimal derivative terms in the operator \cite{Barvinsky:1985an,Saueressig:2011vn}.
These two methods share a common building-block, the generalized heat-kernel traces which contain insertions of covariant derivatives
\be\label{derinsertion}
 \Tr\;\Big[D_{\mu_1}D_{\mu_2}\dots D_{\mu_n}\;H(s)\Big]\, .
\ee
Traces of this type have been studied in \cite{Benedetti:2010nr,Decanini:2005gt,Decanini:2007gj,Anselmi:2007eq} and their 
Seeley-deWitt coefficients can be obtained by a generalization of the early-time expansion \eqref{sdw}.
In particular, they prove very valuable when evaluating traces of operators where the heat-kernel expansion \eqref{hkdef} is not known.

The rest of the paper is structured as follows. In section \ref{main:0} we study traces of the form \eqref{derinsertion}, using the 
deWitt-algorithm for determining their curvature expansion recursively. Implementing this algorithm in a computer algebra system \cite{Mathtensor,Ricci}, we generalize the results of \cite{Benedetti:2010nr,Decanini:2005gt,Decanini:2007gj,Anselmi:2007eq}
to differential operators on a general gauge bundle including an arbitrary endomorphism. As an application of these results, we study the early-time expansion of the traced heat-kernel of the Laplace operator on the space of transversal vector fields in section \ref{main:2}, giving the Seeley-deWitt coefficients up to order $\cR^3$ in table \ref{t.1}. We close by briefly commenting on possible applications of our findings in the context of QFT in section \ref{summary}.  Some technical details have been relegated to four appendices.

\section{The off-diagonal heat-kernel}
\label{main:0}
In this section, we outline how the derivatives of the heat-kernel coefficients \eqref{sdw} can be computed recursively. The main virtue of the off-diagonal method used here is that it allows to
compute operator traces of the general non-minimal form \eqref{derinsertion}, which have a wide range of applications in QFT.

\subsection{Recurrence relation for off-diagonal heat-kernel coefficients}
We assume that our spacetime is a closed Riemannian manifold without boundary and of arbitrary dimension $d$.
The Laplace operator $\D_0$ in the heat-kernel (\ref{hkdef}) is taken to be of general second order form
\be
\begin{split}
\D_0
=\, & - g^{\m\n}\pa_\m\pa_\n + a^\m\pa_\m + b \\
=\, & - g^{\m\n}D_\m D_\n + E \, ,
\end{split}
\ee
where in the second line it is cast into standard notation \cite{Vassilevich:2003xt}, involving a covariant derivative operator $D_\m = \nabla_\m + A_\m$ and an endomorphism $E$.\footnote{This can always be done as long as the manifold is torsionless \cite{Barth:1985cz}.}
The symbol $\D$ is reserved for the Laplacian built from the covariant derivative only $\D=-D^\m D_\m$ (without an endomorphism)
and $\nabla_\m$ is the covariant torsionless spacetime derivative compatible with the metric $g_{\m\n}$.
We define $A_\m$ to be an unspecified connection on an internal bundle over the spacetime manifold and
we assume, without loss of generality, that $E$ is an endomorphism on the same bundle of $A_\m$;
whenever this is not the case it is sufficient to decompose their bundle into the direct sum of their respective bundles.
Further we define the sum of the curvatures of the connection $A_\m$ and the Levi-Civita connection
\be
 F_{\m\n}=\partial_\m A_\n-\partial_\n A_\m +\left[A_\m,A_\n\right] +\left[\nabla_\m,\nabla_\n\right] \, .
\ee
For notational simplicity, all internal indices are suppressed, so that the quantities $A_\m$, $F_{\m\n}$ and $E$ are understood as matrices on the internal space.

The heat-kernel $H(s)$ owes its name to the fact that it fulfills a generalized heat-equation with boundary condition $H(0)=\unit$, where $\unit$ denotes the identity on the internal space.
To arrive at the heat-equation in the form of an ordinary partial differential equation, it is convenient to express $H(s)$ in terms of its matrix elements, called the off-diagonal heat-kernel
\be\label{hkelement}
H(x, y; s) \equiv \langle y | H(s) | x \rangle = \langle y | \e^{-s \D_0} | x \rangle \, ,
\ee
since the basis elements are at different points $x$ and $y$ of the manifold.
The definition (\ref{hkelement}) is equivalently given as an initial value problem of the heat-equation
\be\label{Heateq}
\begin{split}
& \left(\partial_s +\D_{0,x} \right) H(x,y;s) = 0 \, , \qquad 
H(x,y;0) = \d_{x,y} \unit \, .
\end{split}
\ee
The solution of this differential equation has the interpretation of heat propagating on the manifold, according to the operator $\D_{0,x}$, from a source located in $y$. Here $s$ plays the role of a diffusion-time for the process.

The initial value problem (\ref{Heateq}) can be solved explicitely in the simple case of a flat
manifold, where both the connection $A_\m$ and the endomorphism $E$ vanish. In this case the solution of
\eqref{Heateq} is given by 
\be\label{Hflat}
\left.H(x,y;s)\right|_{\rm flat} =
\frac{1}{\left(4\pi s\right)^{d/2}} \e^{-\tfrac{(x-y)^2}{4s}} \, .
\ee
 As an ansatz for the general solution, this expression is modified by introducing a general function $\O(x,y;s)$
\be\label{Hansatz}
H(x,y;s) =
\frac{1}{\left(4\pi s\right)^{d/2}} \e^{-\tfrac{\sigma(x,y)}{2s}}\O(x,y;s) \, .
\ee
Here $\sigma(x,y)$ is half of the squared geodesic distance, called the ``world function'' \cite{DeWitt:1965jb}.
It generalizes the distance measure in (\ref{Hflat}) in a covariant way and satisfies
\be\label{sigmamaster}
\frac{1}{2}\sigma_{;\mu}\sigma^{;\mu} = \sigma
\ee
for arbitrary spacetime points $x$ and $y$.\footnote{We use a semicolon to abbreviate any covariant derivative, $D_\m a \equiv a_{;\m}$. We use the convention that all derivatives are with respect to the coordinate $x$.}
In order to find the partial differential equation satisfied by $\O(x,y;s)$, one substitutes the ansatz \eqref{Hansatz} into the heat-equation \eqref{Heateq} to obtain
\be\label{Omegaeq}
\left(\partial_s +\D_{0,x} \right) H
= \frac{1}{\left(4\pi s\right)^{d/2}} \e^{-\tfrac{\s}{2s}} \left(
- \frac{d}{2s}\O + \frac{1}{2s} \s_{;\m}{}^\m \O + \frac{1}{s} \s_{;\m} \O_{;}{}^\m + \partial_s \O - \O_{;\m}{}^\m + E\O \right) \, .
\ee
The heat-equation is solved if the bracket on the right hand side vanishes. Inserting the early-time expansion of the heat-kernel
\be\label{earlytime}
\O(x,y;s) = \sum_{n\geq 0} s^n A_n(x,y) \, ,
\ee
and requiring that the bracket in \eqref{Omegaeq} vanishes at all orders of $s$ independently, this equation yields the master-equation
\be\label{mastereq}
\left( n - \tfrac{d}{2} + \half \sigma_{;\m}{}^\m \right) A_n + \sigma^{;\m} A_{n;\m} - A_{n-1;\m}{}^\m + E A_{n-1} = 0 \, ,\quad n\ge 0 \, ,
\ee
subject to the initial conditions $A_{-1} = 0$ and $A_0 = \unit$.

Equation \eqref{mastereq} still constitutes a complicate partial differential equation for the coefficients $A_n(x,y)$. In order to solve it, we exploit that
the full off-diagonal heat-kernel coefficients for non-coinciding points can be expressed as the geodesic expansion \cite{Decanini:2005gt}
\be\label{expcoeff}
A_n(x,y) = \sum_{m\ge 0} \frac{(-1)^m}{m!} \,\overline{D_{\mu_1}\dots D_{\mu_m} A_n} \,\sigma^{;\mu_1} \dots \sigma^{;\mu_m} \, ,
\ee
where the overline denotes the coincidence limit of any bi-tensor $C(x,y)$
\be\label{coincidencelimit}
\overline{C(x,y)} \equiv C(x,x)\,.
\ee
Notice that here, it is sufficient to know the quantities $\overline{D_{(\mu_1}\dots D_{\mu_m)} A_n}$ symmetrized in their indices.\footnote{It is important to note that in general $\overline{D_{(\mu_1}\dots D_{\mu_m)} A_n} \neq D_{(\mu_1}\dots D_{\mu_m)} \overline{A_n} $.}
Substituting \eqref{expcoeff} into \eqref{mastereq} then allows to recursively determine the expansion-coefficients \eqref{expcoeff} at coincident points $y \rightarrow x$. Since only covariant expressions are used in its derivation, the $A_n(x,y)$ are given in terms of an expansion in curvature monomials.
To solve \eqref{mastereq} for any $\overline{A_n}$, the coincidence limits of the derivatives of $A_{n-1}$ and $\sigma$ are required.
Comparing powers of the curvatures occurring in these objects, the systematics is easily found and summarized in table \ref{Table.1}.
For example, to compute all coefficients up to the 6-derivative order ($\cR^3$), one needs the coincidence limit of 8 derivatives acting on $\s$,
6 derivatives acting on $A_0$ and so on.
\begin{table}
\begin{center}
\begin{tabular}{|l || c c c c c c c |  } \hline
& ${\cal R}^0$ & ${\cal R}^{1/2}$ & ${\cal R}^1$ & ${\cal R}^{3/2}$ & ${\cal R}^2$ & ${\cal R}^{5/2}$ & ${\cal R}^3$ \\ \hline
& $D^2\sigma$ & $D^3\sigma$ & $D^4\sigma$ & $D^5\sigma$ & $D^6\sigma$ & $D^7\sigma$ & $D^8\sigma$\\
$0$ & $A_0$ & $D^1A_0$ & $D^2A_0$ & $D^3A_0$ & $D^4A_0$ & $D^5A_0$ & $D^6A_0$ \\
$1$ & & & $A_1$ & $D^1A_1$ & $D^2A_1$ & $D^3A_1$ & $D^4A_1$ \\
$2$ & & & & & $A_2$ & $D^1A_2$ & $D^2A_2$ \\
$3$ & & & & & & & $A_3$ \\ \hline
\end{tabular}
\end{center}
\caption{\label{Table.1} Analysis of the terms entering the recursion relation \eqref{mastereq} at coincidence limit (we omit the overline for brevity). In order to compute an entry, one needs to compute every object that is above and on its left. Here ${\cal R}$ counts the number of curvature tensors and of square covariant derivatives. (For example $D^m R^n$ counts ${\cal R}^{n+m/2}$.)}
\end{table}
The $\overline{\sigma_{;\mu_1\dots\mu_n}}$ are found with the help of the defining equation (\ref{sigmamaster}) \cite{Christensen:webtalk}.
With the initial condition $\overline{\s}=0$, it is straightforward to find the first few expressions, by successively applying derivatives to the equation. The coincidence limits up to fifth order in the derivatives are
\be\label{sigmacoinc}
\begin{split}
\overline{\sigma_{;\m}} = 0 \, , & \qquad
\overline{\sigma_{;\m\n}} = g_{\m\n} \, , \\
\overline{\sigma_{;\m\n\r}} = 0 \, , & \qquad
\overline{\sigma_{;\m\n\r\s}} =  - \tfrac{1}{3} \left( R_{\m\r\n\s} + R_{\m\s\n\r} \right) \, , \\
\overline{\sigma_{;\m\n\r\s\a}} = &
- \tfrac{1}{4} \left( R_{\m\n\r\s;\a}+R_{\m\n\r\a;\s}+R_{\m\s\r\n;\a}+R_{\m\s\n\a;\r}+R_{\m\a\n\r;\s}+R_{\m\a\n\s;\r} \right) \, , \\
\end{split}
\ee
and in addition we give
\be\label{sigmacoinc2}
\begin{split}
\overline{\sigma_{;\a\b(\m\n\r\s)}} = &
-\tfrac{12}{5} R_{(\m|\a|\n|\b|;\r\s)}
-\tfrac{4}{5} R_{(\m\n|\a|}{}^\g R_{\r|\b|\s)\g}
-\tfrac{4}{5} R_{\g(\m\n|\a|} R_{\r\s)\b}{}^\g
+\tfrac{8}{15} R_{\g(\m\n|\a|} R_{\r|\b|\s)}{}^\g \\
& +\tfrac{16}{45} R_{\g(\m\n\r} R_{\s)\a\b}{}^\g
-\tfrac{8}{15} R_{\g(\m\n\r} R_{\s)\b\a}{}^\g
+\tfrac{4}{9} R_{\g(\m\n\r} R_{\s)}{}^\g{}_{\a\b} \, .
\end{split}
\ee
These objects can be found as follows. After taking $n$ derivatives of (\ref{sigmamaster}), terms appearing with $n+1$ derivatives vanish in the coincidence limit because $\overline{\sigma_{;\m}}$ does. Inserting all lower order results and commuting the $n$ indices in the remaining terms into a unique order immediately reveals the result. The method is conceptually straightforward and requires only simple algebra, yet the higher order terms increase in size quickly and render a by-hand-computation unfeasible.

Equation~\eqref{mastereq} is solved for the quantities $\overline{D_{\mu_1}\dots D_{\mu_m} A_n}$ for any $n, m \ge 0$ in the same way.
One can obtain an infinite set of algebraic equations for these by applying derivatives and taking the coincidence limit of \eqref{mastereq}.
These can be solved for recursively, substituting all lower order objects to find the next order. Once all required derivatives are found for some $n$, one can proceed to $n+1$ until all ingredients to the heat-kernel expansion to a desired order are found.
The results are listed in the next subsection.

Notably, the recursive relation for the heat-kernel coefficients (\ref{mastereq}) becomes \emph{independent} of the spacetime dimension once the coincidence limit is taken, since $\overline{\s_{;\m}{}^\m}=d$ cancels the multiplicative $d$ in (\ref{mastereq}). This is an important observation because it implies that the heat-kernel coefficients of Laplace-type operators cannot  depend on the dimension explicitly. We stress that this property  does not hold for more general operators like \eqref{nonminimal_operator}.

\subsection{Heat-kernel coefficients on a general vector bundle}
\label{main:1}

Starting from the discussion of the previous subsection, it is straightforward to implement the recursive equations \eqref{sigmamaster} and  \eqref{mastereq} in a computer algebra software \cite{Mathtensor,Ricci} to find the $\overline{A_{n;\m_1..\m_m}}$ explicitly.
Up to second order in the curvatures, the coincidence limit of the coefficients and their derivatives are found as
\begin{eqnarray}\label{coefficients}
\overline{A_0} &=& 1 \,,\nonumber\\
\overline{D_\mu A_0} &=& 0 \,,\nonumber\\
\overline{D_{(\nu} D_{\mu)} A_0} &=& \frac{1}{6}R_{\nu\mu} \,,\nonumber\\
\overline{D_{(\alpha} D_\nu D_{\mu)} A_0} &=& \frac{1}{4} R_{(\nu\mu;\a)} \,,\nonumber\\
\overline{D_{(\beta} D_\alpha D_\nu D_{\mu)} A_0} &=& \frac{3}{10} R_{(\nu\mu;\a\b)} +\frac{1}{12} R_{(\beta\alpha}R_{\nu\mu)} +\frac{1}{15} R_{\gamma(\beta|\delta|\alpha} R^{\gamma}{}_{\nu}{}^{\delta}{}_{\mu)} \,,\nonumber\\
\overline{A_1} &=& -E+\frac{1}{6} R \,,\nonumber\\
\overline{D_\mu A_1} &=& -\frac{1}{2} E_{;\m} - \frac{1}{6} F_{\nu\mu;}{}^\n + \frac{1}{12} R_{;\m} \,,\\
\overline{D_{(\nu} D_{\mu)} A_1} &=&
-\frac{1}{3} E_{;(\m\n)} - \frac{1}{6} R_{\mu\nu} E -\frac{1}{6} F_{\alpha(\mu;}{}^\a{}_{\n)} +\frac{1}{6} F_{\alpha(\nu}F^\alpha{}_{\mu)} \nonumber\\
&& +\frac{1}{20} R_{;(\m\n)} -\frac{1}{60} \D R_{\nu\mu} +\frac{1}{36} R R_{\nu\mu} \nonumber\\
&& -\frac{1}{45} R_{\nu\alpha}R^\alpha{}_\mu +\frac{1}{90} R_{\alpha\beta} R^\alpha{}_\nu{}^\beta{}_\mu +\frac{1}{90} R^{\alpha\beta\gamma}{}_\nu R_{\alpha\beta\gamma\mu} \,,\nonumber\\
\overline{A_2} &=&
\frac{1}{6} \D E +\frac{1}{2} E^2 - \frac{1}{6} R E +\frac{1}{12} F_{\mu\nu}F^{\mu\nu} \nonumber\\
&& -\frac{1}{30} \D R +\frac{1}{72} R^2 -\frac{1}{180} R_{\mu\nu}R^{\mu\nu} +\frac{1}{180} R_{\mu\nu\alpha\beta}R^{\mu\nu\alpha\beta} \,. \nonumber
\end{eqnarray}
We also computed all coefficients up to third order in the curvatures (${\cal R}^3$). Here we state only the explicit result
\begin{eqnarray}\label{A3}
\overline{A_3} &=&
-\frac{1}{6} E^3
-\frac{1}{12} (\D E) E
+\frac{1}{12} E_{;\m} E_;{}^\m
-\frac{1}{12} E (\D E)
-\frac{1}{60} (\D \D E) \nn\\&&
+\frac{1}{60} E_{;\n} F^{\m\n}{}_{;\m}
-\frac{1}{60} F^{\m\n}{}_{;\m} E_{;\n}
-\frac{1}{20} E F_{\m\n} F^{\m\n}
-\frac{1}{90} F_{\m\n} E F^{\m\n}
-\frac{1}{45} F_{\m\n} F^{\m\n} E \nn\\&&
-\frac{1}{90} (\D F_{\m\n}) F^{\m\n}
+\frac{1}{45} F_{\m\n;\r} F^{\m\n;\r}
-\frac{1}{180} F_{\m\n} (\D F^{\m\n}) \nn\\&&
+\frac{1}{180} F^\m{}_{\n;\m} F^{\r\n}{}_{;\r}
+\frac{1}{45} F^{\m\n} F_\m{}^\r{}_{;\r\n}
+\frac{1}{90} F^{\m\r}{}_{;\r\n} F_\m{}^\n
-\frac{1}{90} F_{\m\n} F^{\n\r} F^\m{}_\r \nn\\&&
+\frac{1}{12} E^2 R
+\frac{1}{36} (\D E) R
-\frac{1}{30} E_{;\m} R_;{}^\m
+\frac{1}{30} E (\D R)
-\frac{1}{90} E_{;\m\n} R^{\m\n} \nn\\&&
-\frac{1}{72} E R^2
+\frac{1}{180} E R_{\m\n} R^{\m\n}
-\frac{1}{180} E R_{\m\n\r\s} R^{\m\n\r\s} \\&&
+\frac{1}{72} F_{\m\n} F^{\m\n} R
+\frac{1}{30} F_{\m\n} F^{\m\r} R^\n{}_\r
-\frac{1}{180} F_{\m\n} F_{\r\s} R^{\m\n\r\s} \nn\\&&
+\frac{1}{280} (\D \D R)
-\frac{1}{280} R (\D R)
+\frac{17}{5040} R_{;\m} R_;{}^\m
+\frac{1}{420} R_{;\m\n} R^{\m\n}
+\frac{1}{630} R_{\m\n} (\D R^{\m\n}) \nn\\&&
-\frac{1}{2520} R_{\m\n;\r} R^{\m\n;\r}
-\frac{1}{1260} R_{\m\n;\r} R^{\n\r;\m}
-\frac{1}{420} R_{\m\n\r\s} (\D R^{\m\n\r\s})
+\frac{1}{560} R_{\m\n\r\s;\l} R^{\m\n\r\s;\l} \nn\\&&
+\frac{1}{1296} R^3
-\frac{1}{1080} R R_{\m\n} R^{\m\n}
+\frac{1}{5670} R_{\m\n} R^{\m\r} R^\n{}_\r \nn\\&&
-\frac{1}{1890} R_{\m\n} R_{\r\s} R^{\m\r\n\s}
+\frac{1}{1080} R R_{\m\n\r\s} R^{\m\n\r\s}
-\frac{1}{945} R_{\m\n} R_{\m\r\s\t} R^\n{}_{\r\s\t} \nn\\&&
+\frac{1}{567} R^\m{}_\n{}^\r{}_\s R_{\m\a\r\b} R^{\n\a\s\b}
+\frac{11}{11340} R^{\m\n}{}_{\r\s} R_{\m\n\a\b} R^{\r\s\a\b} \nn \, .
\end{eqnarray}
Some additional coefficients at third order in the curvatures are given in appendix \ref{App.3}.
The ones that are not given are too large to be conveniently written.
This formula generalizes previous computations of $\overline{A_3}$ with traced internal space indices \cite{Vassilevich:2003xt}, and in the limit of trivial connection ($F_{\mu\nu}=0$) \cite{Decanini:2005gt}.
Note that in the above expressions, a full covariant derivative $D$ is identical to the covariant spacetime derivative $\nabla$ when it acts on the spacetime curvature tensors $R$.

\subsection{Evaluating traces with non-minimal differential operator insertions}
\label{sect:2.3}

The derivatives of the heat-kernel coefficients can be used to evaluate operator traces with non-minimal derivative insertions. The trace of the heat-kernel defined in terms of its matrix elements (\ref{hkelement}) is
\be
\begin{split}
\Tr[\e^{-s\D_0}]
=\, & \tr \int d^d x \sqrt{g}\; \left<x\right|\e^{-s\D_0}\left|x\right> \\
=\, & \tr \int d^d x \sqrt{g}\; H(x,x;s) \, ,
\end{split}
\ee
where ``$\tr$'' is the trace in the internal space. This expression generalizes to the case of arbitrary insertions of uncontracted covariant derivatives like
\be\label{nonmintrace}
\begin{split}
\Tr [D_{\m_1} \dots D_{\m_n} \e^{-s\D_0}]
=\, & \tr \int d^d x \sqrt{g}\; \left<x\right| D_{\m_1} \dots D_{\m_n} \e^{-s\D_0}\left|x\right> \\
=\, & \tr \int d^d x \sqrt{g}\; H_{\m_1\dots\m_n}(x,s) \, ,
\end{split}
\ee
where we introduced the abbreviations for the coincidence limits
\be
H_{\m_1\dots\m_n}(x,s) \= \overline{D_{\m_1} \dots D_{\m_n} H(x,y;s)} = \left<x\right| D_{\m_1} \dots D_{\m_n} \e^{-s\D_0} \left|x\right> \, .
\ee
The derivatives acting on $H(x,y;s)$ can be applied explicitly to the definition (\ref{Hansatz}), in order to express $H_{\m_1\dots\m_n}(x,s)$ in terms of the heat-kernel coefficients (\ref{coefficients}). Hereby the derivatives will act on both $\s(x,y)$ and $\O(x,y;s)$, where the latter is given by the heat-kernel coefficients via (\ref{earlytime}).
We give the result for the first six of these matrix elements with symmetrized derivatives:
\be\label{hsymder}
\begin{split}
H(x,s) =&
\left(4 \pi s\right)^{-d/2} \sum_{n\ge 0} s^n \overline{A_n} \\
H_{\m}(x,s) =&
\left(4 \pi s\right)^{-d/2} \sum_{n\ge 0} s^n \overline{D_\m A_n} \\
H_{(\m\n)}(x,s) =&
\left(4 \pi s\right)^{-d/2} \sum_{n\ge 0} s^{n-1} \Bigl(
 -\frac{1}{2} g_{\m\n} \overline{A_n}
 +\overline{D_{(\m}D_{\n)} A_{n-1}}
\Bigr) \\
H_{(\m\n\r)}(x,s) =&
\left(4 \pi s\right)^{-d/2} \sum_{n\ge 0} s^{n-1} \Bigl(
 -\frac{3}{2} g_{(\r\n} \overline{D_{\m)} A_n}
 +\overline{D_{(\r}D_\n D_{\m)} A_{n-1}}
\Bigr) \\
H_{(\m\n\r\l)}(x,s) =&
\left(4 \pi s\right)^{-d/2} \sum_{n\ge 0} s^{n-2} \Bigl(
 \frac{3}{4} g_{(\l\r}g_{\n\m)} \overline{A_n}
 -3 g_{(\l\r} \overline{D_\n D_{\m)} A_{n-1}} \\
&
 +\overline{D_{(\l} D_\r D_\n D_{\m)} A_{n-2}}
\Bigr) \\
H_{(\m\n\r\l\a)}(x,s) =&
\left(4 \pi s\right)^{-d/2} \sum_{n\ge 0} s^{n-2} \Bigl(
 \frac{15}{4} g_{(\a\l}g_{\r\n} \overline{D_{\m)}A_n} \\
&
 -5 g_{(\a\l} \overline{D_\r D_\n D_{\m)} A_{n-1}}
 +\overline{D_{(\a}D_\l D_\r D_\n D_{\m)} A_{n-2}}
\Bigr) \\
H_{(\m\n\r\l\a\b)}(x,s) =&
\left(4 \pi s\right)^{-d/2} \sum_{n\ge 0} s^{n-3} \Bigl(
 -\frac{15}{8} g_{(\b\a}g_{\l\r}g_{\n\m)} \overline{A_n} \\
&
 +\frac{45}{4} g_{(\b\a} g_{\l\r} \overline{D_\n D_{\m)} A_{n-1}}
 -\frac{15}{2} g_{(\b\a} \overline{D_\l D_\r D_\n D_{\m)} A_{n-2}} \\
&
 +\overline{D_{(\b} D_\a D_\l D_\r D_\n D_{\m)} A_{n-3}}
\Bigr)
\end{split}
\ee
In these expressions, the coincidence limits of derivatives of $\s$ (\ref{sigmacoinc}) have been inserted, and the sums are understood with the boundary conditions $\overline{A_{-1}}=\overline{A_{-2}}=\dots=0$. The general unsymmetrized formulas can always be recovered from the symmetrized ones by commuting the derivatives.
For example we have
\be
 \begin{split}
 H_{\m\n}(x,s)
 =& H_{(\m\n)}(x,s) + \left(4 \pi s\right)^{-d/2} \sum_{n\ge 0} s^{n-1} \overline{D_{[\m}D_{\n]} A_{n-1}}\\
 =& H_{(\m\n)}(x,s) + \left(4 \pi s\right)^{-d/2} \sum_{n\ge 0} s^{n-1} F_{\m\n} \overline{A_{n-1}}\, ,
 \end{split}
\ee
where the commutator in the second term becomes the curvature tensor. With the explicit knowledge of the expansion coefficients (\ref{coefficients}), all traces of the form (\ref{nonmintrace}) are given.

The relevance of these traces lies in the fact that quite general expressions involving the Laplacian operator and uncontracted derivatives can be computed by use of (\ref{hsymder}).
For example
\be\label{non-minimal-trace}
\Tr[D_{\m_1}\dots D_{\m_n} f\left(\Delta_0\right)] =
\tr \int d^d x \sqrt{g}\; \int ds\; \tilde{f}\left(s\right) \overline{D_{\m_1} \dots D_{\m_n} H(x,y;s)} \, ,
\ee
where a general function of $\D_0$ is written as a Laplace-transform $f\left(\D_0\right) = \int ds\; \tilde{f}\left(s\right) \e^{-s\D_0}$. This technique has been successfully applied to sophisticated traces appearing in functional renormalization group calculations \cite{Benedetti:2010nr,Saueressig:2011vn,Groh:2010ta,Codello:2011yf}, where it allowed to generalize the computations to operator traces which are not accessible by standard heat-kernel technique.

Before closing this section, it is worth noting that a non-local generalization of (\ref{hsymder}) appeared in the literature \cite{Gusev}.
Although the result covers only to the first order in the curvatures only, it shows that the generalization of (\ref{hsymder}) to a non-local heat-kernel expansion like (\ref{bv}) is possible.
Notably, such a generalization
would give many insights to important physical problems, like the computation of radiative currents.

\subsection{Tensor fields on a curved background}
\label{ssect:2.4}
In this subsection, we show how to obtain the heat-kernel for the special case of a Laplace operator acting on scalar-, vector- and symmetric tensor fields from the general formulas \eqref{coefficients}, \eqref{A3} and \eqref{hsymder}.
In this case we set $E = 0$ and choose $A_\mu=0$, such that the field strength becomes
$F_{\m\n} = [\nabla_\m, \nabla_\n]$ acting on scalars, vectors and symmetric tensors, respectively.

From the commutators acting on the field spaces of interest we compute
\be\label{commutators}
\begin{split}
 \left[\nabla_\m,\nabla_\n\right] \phi
 &=
 0\, ,\\
 \left[\nabla_\m,\nabla_\n\right] v^{\a}
 &=
 R_{\m\n}{}^\a{}_\b v^\b\, , \\
 \left[\nabla_\m,\nabla_\n\right] h^{\a\b}
 &=
 2R_{\m\n}{}^{(\a}{}_{(\gamma} \delta{}^{\b)}_{\delta)} h^{\gamma\delta}\, ,
\end{split}
\ee
where $\phi$, $v^\alpha$ and $h^{\alpha\beta}$ are test scalar-, vector- and symmetric tensor fields respectively.
From \eqref{commutators} the following relations between field strengths and Riemann curvatures are implied
\be\label{field_strenghts}
\begin{split}
 \left.F_{\m\n}\right|_{\rm scalar}
 &=
 0\, ,\\
 \left.F_{\m\n}{}^\a{}_\b\right|_{\rm vector}
 &=
 R_{\m\n}{}^\a{}_\b \, , \\
 \left.F_{\m\n}{}^{\a\b}{}_{\gamma\delta}\right|_{\rm tensor}
 &=
 2R_{\m\n}{}^{(\a}{}_{(\gamma} \delta{}^{\b)}_{\delta)}\, .
\end{split}
\ee
Once \eqref{field_strenghts} are computed, we can substitute them into \eqref{hsymder} and obtain the heat-kernel expansion for the special cases.
The diagonal heat-kernel expansion and its coefficients are defined by
\be\label{earlytimeexp}
{\rm Tr}_r\left[\e^{-s\Delta}\right] = \frac{1}{(4\pi s)^{d/2}} \int d^dx \sqrt{g} \left[ c^0 \cR^0 + s \, c^1 \, \cR^1 + s^2 \sum_{i=1}^3 c_i^2 \, \cR_i^2 + s^3 \sum_{i=1}^{10} c_i^3 \, \cR^3_i\right]
\ee
where $r$ labels the three tensor space of interest and we adopted the basis \eqref{basis}.
The coefficients of \eqref{earlytimeexp} are listed in table \ref{t.1} together with the results of the next section.

\section{Seeley-deWitt expansion for transverse vector fields}
\label{main:2}

In this section we compute the early-time expansion of the heat-kernel resulting from a projected Laplace operator acting on transverse vector fields, up to third order in the curvature. To simplify matters, we will consider abelian vector fields setting the connection $A_\m = 0$ for the remainder of this section. In this case $D_\m = \nabla_\m$ and we use the former notation throughout. The final result \eqref{totalresult} holds for compact manifolds without boundary. Throughout the computation we will make extensive use of the off-diagonal heat-kernel coefficients derived in the previous section.

\subsection{Transversal vectors}
\label{sect:3.1}
In Yang-Mills theory on flat space it is a common practice to decompose the fluctuations of the vector field $v^\mu$ into its transversal and longitudinal parts \cite{PStextbook}. In Lorentz-gauge the gauge-transformations solely act on the transversal part, so that the latter constitutes a gauge degree of freedom of the theory. Following \cite{Reuter:1993kw}, the decomposition
\be
v^\mu = \xi^\mu + \p^\mu \phi \, , \qquad \partial_\m \xi^\mu = 0
\ee
can be implemented by introducing the projection operators
\be\label{flatspaceprojectors}
\Pi_{\rm L}{}_\m{}^\n \equiv \frac{\p_\m \p^\n}{\p^2} \, , \qquad \Pi_{\rm T}{}_\m{}^\n \equiv \delta_\m^\n - \frac{\p_\m \p^\n}{\p^2} 
\ee
which satisfy $ \Pi^2 =  \Pi$ and $ \Pi_{\rm L}{}_\m{}^\n \p_\n = \p_\m$, $\p^\m  \Pi_{\rm L}{}_\m{}^\n  = \p^\n$. In terms of these
\be
\p^\mu \phi = \Pi_{\rm L}{}_\m{}^\n \, v_\n \, , \qquad \xi^\mu =  \Pi_{\rm T}{}_\m{}^\n \, v_\n \, .
\ee

In the context of gravity or quantum field theory on a curved spacetime, it is desirable to extend this decomposition to a general Riemannian (background) manifold.
In this case, one has the local decomposition \cite{York:1973ia}
\be\label{Tdec}
\begin{split}
 v^\mu =& \xi^\mu + D^\mu \phi \, , \qquad D_\mu \xi^\mu \equiv 0 \, .
\end{split}
\ee
This split is unique up to a constant shift in $\phi$, which constitutes the zero-mode of $D_\m$. The generalization
of the projectors \eqref{flatspaceprojectors} implementing this decomposition is straightforward
\be\label{projectors}
\begin{split}
\P_{\rm L}{}_{\m}{}^{\n} \equiv -D_\m \frac{1}{\D} D^\n \, , \qquad \P_{\rm T}{}_{\m}{}^{\n} \equiv \delta_{\m}^{\n}+D_\m \frac{1}{\D} D^\n \, .
\end{split}
\ee
They project $v^\m$ to its irreducible components
\be
\begin{split}
 \P_{\rm T}{}_{\m}{}^{\n} v_\n = \xi_\m \, , \qquad
 \P_{\rm L}{}_{\m}{}^{\n} v_\n = D_\m \phi \, .
\end{split}
\ee

In order to construct a well-defined operator trace, it is crucial to observe that the eigenvalue equation for the
standard Laplace operator is, in general, incompatible with the transverse condition. This can be
seen as follows: Suppose that $\xi_i^\m$ is an eigenfunction of $\Delta$ on the space of transverse vector fields
\be\label{EVeq}
\D \xi_i^\m = \lambda_i \xi_i^\m \, , \qquad \lambda_i > 0 \, , \; D_\m \xi_i^\m = 0 \, .
\ee
Applying $D_\m$ to the left-hand-side of this equation gives
\be
 D_\m \D \xi_i^\m =  \left[ D_\m , \D \right] \xi_i^\m = -D^\m R_{\m\n} \xi_i^\n
\ee
where we used the transverse condition together with \eqref{eqB6}. The divergence of the right-hand-side of \eqref{EVeq} vanishes identically, however, so that we obtain the condition
\be\label{Decomposition}
D^\m R_{\m\n} \xi_i^\n = 0 \, .
\ee
On a general Riemannian manifold there is no reason that this identity holds. Thus on a Riemannian manifold the space of eigenfunctions of $\D$ does not decompose into the direct sum of transverse and longitudinal vector fields.

An important special case are the Einstein manifolds discussed in appendix \ref{Einstein}. These have the special property that $R_{\m\n} \propto g_{\m\n}$, so that \eqref{Decomposition} is satisfied. In other words the Einstein-condition ensures $[\D, \P_{\rm T}{}^\m{}_\n] v^\n = 0$ which can be verified easily with the use of the equations (\ref{commfuncES}).
Thus, for this special case, the space of eigenfunctions of $\D$ \emph{does decompose} into the direct sum of transverse and longitudinal vector fields.

In order to give a well-defined meaning to the heat-kernel on the restricted set of $\xi^\mu$ configurations satisfying the constraint $D_\mu \xi^\mu=0$, we define the projected Laplace operator
\be\label{1TLap}
\tilde{\D}{}^\m{}_\n \equiv \P_{\rm T}{}^\m{}_\a \Delta \P_{\rm T}{}^\a{}_\n \, .
\ee
The projectors ensure, that $\tilde{\D}{}^\m{}_\n$ propagates the transversal modes only. Moreover, even for the case of a general Riemannian manifold, all eigenfunctions of $\tilde{\D}{}^\m{}_\n$ are transversal vector fields by construction.  The projectors entering into the definition \eqref{1TLap} turn $\tilde{\D}{}^\m{}_\n$ into a non-local, pseudo-differential operator, however. Thus it is a priori unclear to which extend the standard results for the Seeley-deWitt expansion carry over to the case of vector fields satisfying differential constraints. This will be investigated in the next subsections.

\subsection{Structural aspects of the heat-trace}
\label{main:2.2}
Based on the previous discussion, we are now in the position to give a well-defined meaning to the heat-trace on the space of transversal vector fields, Tr$_{\rm 1T} \e^{-s \D}$, as the traced heat-kernel of the projected Laplacian $\tilde{\D}$:
\be\label{1Ttrace}
\begin{split}
S_{\rm 1T} \;\equiv\; & {\rm Tr}_1 \; \Pi_{\rm T} \, \e^{-s \Pi_{\rm T} \Delta \Pi_{\rm T}} \, .
\end{split}
\ee
The additional projector in front of the exponential serves the purpose to remove the contribution of the longitudinal vector modes from the zeroth order of the exponential series. Our aim is to compute the early-time expansion of $S_{\rm 1T}$ up to third order in the curvature. In this context, it is important to note that since $\tilde{\D}$ is a non-local operator and in particular not of generalized Laplace-type, the results of the previous section cannot be applied straightforwardly.

The explicit computation of $S_{\rm 1T}$ starts from expanding the exponential appearing in \eqref{1Ttrace}
in a power series
\be\label{powexp}
S_{\rm 1T} = \sum_{n\geq 0} \frac{(-s)^n}{n!} \, {\rm Tr}_1 \, \Pi_{\rm T} \, \left(\Delta\Pi_{\rm T}\right)^n \, .
\ee
Here we used the idempotency of $\P_{\rm T}$ to remove the right set of projectors appearing in $\left(\Pi_{\rm T}\Delta\Pi_{\rm T}\right)^n$. Subsequently, we apply the commutation relations between the projectors and $\Delta$ to collect all powers of the (unprojected) Laplacian. Resumming the power series for each type of commutator, eq.\ \eqref{powexp} can be expanded systematically. To lowest order in the commutator this expansion has the form
\be\label{curvexp}
S_{\rm 1T} = 
{\rm Tr}_1 \, \Big[ \Pi_{\rm T} \, \e^{-s  \Delta} \Big]
- s \, {\rm Tr}_1 \, \Big[  \Pi_{\rm T} \, \left[ \, \Delta \, , \, \Pi_{\rm T} \, \right] \, \e^{-s  \Delta} \Big] + \ldots
\ee
and the dots are symbolic for higher order commutators. Each commutator carries fixed power of (derivatives) of the curvature tensor.
For instance, the lowest order commutator is
\be\label{firstcom}
\Big[ \D, \P_{\rm T}{}_\m{}^\n \Big] v_\n = \Big( \P_{\rm L}{}_\m{}^\a R_\a{}^\n - R_\m{}^\a \P_{\rm L}{}_\a{}^\n \Big) v_\n \, .
\ee
The higher order commutators then contain more powers of the curvature tensor or its derivatives.
Thus eq.\ \eqref{curvexp} constitutes 
a curvature expansion which allows to evaluate $S_{\rm 1T}$ up to a specified order in $\cR$.

In order to apply the off-diagonal heat-kernel technique to this curvature expansion, we next have to deal with 
the inverse powers of the Laplacians appearing in the projectors and their commutators. Here we employ the Schwinger-trick
\be\label{tInt0}
\frac{1}{\D^n} = \int_0^\infty \frac{t^{n-1}}{(n-1)!} \e^{-t\D} \, dt \, .
\ee
By combining all exponentials including the Laplace operator by applying further commutator relations if necessary, the traces in \eqref{curvexp}
can be cast into the form
\be\label{tInt1}
{\rm Tr}_1 \, \Big[\cO \; \frac{1}{\D^n} \, \e^{-s \Delta} \Big] =
\int_0^\infty \frac{t^{n-1}}{(n-1)!} \, {\rm Tr}_1 \, \Big[\cO \; \e^{-(s+t) \Delta} \Big] dt \, ,
\ee
where $\cO = \cR D_{\m_1} \ldots D_{\m_n}$ constitute non-minimal operator insertions of the form \eqref{nonmintrace}. 
Thus the operator traces on the right-hand-side can  be evaluated via the off-diagonal heat-kernel technique of section \ref{sect:2.3}. 

The crucial new feature appearing in eq.\ \eqref{tInt1} is the auxiliary $t$-integral, which can be traced back to the non-locality
in the projection operators. Upon inserting the $H$-functions \eqref{hsymder}, these integrals assume the generic form
\be\label{tInt2}
I(n,k-\tfrac{d}{2}) \; := \;
\int_0^\infty \frac{t^{n-1}}{(n-1)!} (s+t)^{k-d/2} dt \;=\; \frac{\Gamma(\tfrac{d}{2}-k-n)}{\Gamma(\tfrac{d}{2}-k)} \, s^{n+k-\tfrac{d}{2}} \, .
\ee
The gamma functions appearing here show the typical form of an IR divergence as seen within dimensional regularization.
Indeed the r.h.s. of \eqref{tInt2} becomes singular for $n+k \ge d/2$, that is, if the order of the heat-kernel expansion $\cR^{n+k}$ exceeds $d/2$.
One would thus expect that the coefficients in the Seeley-deWitt expansion of $S_{\rm 1T}$ will develop divergences at $\cO(\cR^{d/2})$. With respect to the explicit computation in the next subsection, we will use \eqref{tInt2} to regularize the Seeley-deWitt coefficients by analytically continuing in the spacetime dimension $d$, assuming that it is sufficiently large for all integrals to converge.

Before closing our discussion, we remark that it is necessary to keep the spacetime dimension unspecified, so that no terms are lost in the traced heat-equation
\be
\begin{split}
{\rm Tr} \Big[ \D H(s) \Big]
& = -\tfrac{d}{ds}\; {\rm Tr} \Big[ H(s) \Big] \\
& = (4 \pi)^{-d/2} \sum_{n \ge 0} (\tfrac{d}{2}-n) s^{n-1-d/2} A_n \, .
\end{split}
\ee
Integrating this relation should reproduce the standard heat-kernel
\be
{\rm Tr} \big[H(s)\big]=\int_0^\infty {\rm Tr} \big[\D H(s+t)\big] dt \, ,
\ee
which requires that none of the factors $(\tfrac{d}{2}-n)$ vanishes in the above formula.

\subsection{Evaluation of the heat-trace}
\label{main:2.3}
After discussing the structural aspects related to the evaluation of \eqref{1Ttrace}, we proceed carrying out the explicit computation including all terms up to third order in the curvature expansion.
Our first task is to complete the curvature expansion \eqref{curvexp} by including all (multi-)commutators which give rise to contributions up to order $\cR^3$. To find the coefficients multiplying the basis monomials \eqref{basis}, it is sufficient to keep track of all operator insertions which contain up to three powers of the curvature or terms where two covariant derivatives act on two curvature tensors. Terms where three or four covariant derivatives act on one curvature tensor are, in principle also of order $\cR^3$, but give rise to surface-terms only and may thus be discarded. From \eqref{firstcom} we learn that $[\D, \P_{\rm T}] \propto \cR$. Taking an additional commutator $[\D, [\D, \P_{\rm T}]]$ then gives terms with two powers of the curvature or one covariant derivative acting on $R_{\m\n}$. From this systematics we conclude that the highest multi-commutator contributing to $\cO(\cR^3)$ is the quadruple commutator $\left[\Delta, \left[\Delta, \left[\Delta, \left[\Delta, \Pi_{\rm T} \right] \right] \right] \right]$. In addition one has to keep all products of lower order commutators, that satisfy the above criteria. Starting from \eqref{powexp} and working out the combinatorics,  the curvature expansion of $S_{\rm 1T}$ up to  $\cO(\cR^3)$ is given by
\be\label{1Ttrace_expanded}
S_{\rm 1T} \simeq \sum_{n=0}^4 \frac{1}{n!} \, (-s)^n \, S_{\rm 1T}^{(n)} \, , 
\ee
where
\be\label{S1Tn}
\begin{split}
S_{\rm 1T}^{(0)} = & \,
{\rm Tr}_1 \, \Pi_{\rm T} \, \e^{-s  \Delta} \\
S_{\rm 1T}^{(1)} = & \, {\rm Tr}_1 \, \Pi_{\rm T} \, \left[ \, \Delta \, , \, \Pi_{\rm T} \, \right] \, \e^{-s  \Delta} \\
S_{\rm 1T}^{(2)} = & \, {\rm Tr}_1 \, \Pi_{\rm T} \, \left( \left[ \, \Delta \, , \, \Pi_{\rm T} \, \right]^2 + \left[ \, \Delta \, , \, \left[ \, \Delta \, , \, \Pi_{\rm T} \, \right] \right] \right) \, \e^{-s  \Delta} \\
S_{\rm 1T}^{(3)} = & \, {\rm Tr}_1 \, \Pi_{\rm T} \, \bigg(
2 \left[ \, \Delta \, , \, \Pi_{\rm T} \, \right] \left[ \, \Delta \, , \, \left[ \, \Delta \, , \, \Pi_{\rm T} \, \right] \right] + \left[ \, \Delta \, , \, \left[ \, \Delta \, , \, \Pi_{\rm T} \, \right] \right] \left[ \, \Delta \, , \, \Pi_{\rm T} \, \right] \\
& \qquad \qquad \qquad  +  \left[ \, \Delta \, , \, \left[ \, \Delta \, , \, \left[ \, \Delta \, , \, \Pi_{\rm T} \, \right] \right] \right] + \left[ \, \Delta \, , \, \Pi_{\rm T} \, \right]^3
\bigg) \, \e^{-s  \Delta} \\
S_{\rm 1T}^{(4)} = & \, {\rm Tr}_1 \, \Pi_{\rm T} \, \bigg(
3 \left[ \, \Delta \, , \, \Pi_{\rm T} \, \right] \left[ \, \Delta \, , \, \left[ \, \Delta \, , \, \left[ \, \Delta \, , \, \Pi_{\rm T} \, \right] \right] \right]
+ 3 \left[ \, \Delta \, , \, \left[ \, \Delta \, , \, \Pi_{\rm T} \, \right] \right]^2 \\
& \qquad \qquad \qquad + \left[ \, \Delta \, , \, \left[ \, \Delta \, , \, \left[ \, \Delta \, , \, \Pi_{\rm T} \, \right] \right] \right] \left[ \, \Delta \, , \, \Pi_{\rm T} \, \right]
+ \left[ \Delta \, , \, \left[ \, \Delta \, , \, \left[ \, \Delta \, , \, \left[ \, \Delta \, , \, \Pi_{\rm T} \, \right] \right] \right] \right]
\bigg) \, \e^{-s  \Delta} \, ,
\end{split}
\ee
and $\simeq$ indicates that the right-hand-side is actually a curvature expansion.

The actual evaluation of the $S_{\rm 1T}^{(n)}$ can then be carried out along the lines of the previous subsection. Since the computation is rather lengthy and not very illuminating, we relegate these technical details to appendix \ref{App.4}. The final result for $S_{\rm 1T}$ is obtained by substituting the intermediate results \eqref{S1T0final}, \eqref{S1T1final}, \eqref{S1T2final}, \eqref{S1T3final} and \eqref{S1T4final} into \eqref{1Ttrace_expanded}. In terms of the basis \eqref{basis} it takes the form
\be\label{totalresult}
S_{\rm 1T} = \frac{1}{(4\pi s)^{d/2}} \int d^dx \sqrt{g} \left[ c^0 \cR^0 + s \, c^1 \, \cR^1 + s^2 \sum_{i=1}^3 c_i^2 \, \cR_i^2 + s^3 \sum_{i=1}^{10} c_i^3 \, \cR^3_i\right]
\ee
with the coefficients $c^n_m$ given in the last column of table \ref{t.1}. Eq.\ \eqref{totalresult} constitutes the main result of this section.

Commenting on the result for the $S_{\rm 1T}$ trace, first we notice that it has poles in even spacetime dimensions.
Their origin can be traced back to the $t$-integrals in \eqref{tInt2}, that are finite for sufficiently large $d$ only.
We observe that the coefficients at order ${\cal R}^n$ diverge for $n > d/2$. Thus, the poles occur in one order higher then estimated via power counting, so the highest $d$ poles at each order in the curvature cancel, allowing us to analytically continue the results to lower dimensionality. Unfortunately, this phenomenon does not occur for all the poles and we are left with the bounds $d>2$ and $d>4$ for the orders ${\cal R}^2$ and ${\cal R}^3$. A more detailed discussion on these poles will be given in the next subsection.
Notably, all results can be safely continued to any odd dimensionality.
This is owed to the fact that, for odd $d$, the gamma functions in \eqref{tInt2} never become singular.

The coefficients of the transverse vector trace, given in table \ref{t.1}, do not factorize into the difference of those of the vector and the scalar trace.
This is what one would naively expect by simply counting the degrees of freedom, but is obviously wrong because of the presence of ``interaction terms'' between transverse and longitudinal modes.
Interestingly, however, there is a partial decoupling of the degrees of freedom at any order in the curvatures. In fact, it is always possible to write any coefficient of the transverse trace as the aforementioned difference modulo a correction:
\begin{eqnarray}
 c^j_i\left({\rm transverse}\right) &=& c^j_i\left({\rm vector}\right)-c^j_i\left({\rm scalar}\right)+Q_1\left(d\right)/Q_2\left(d\right)\, ,
\end{eqnarray}
where $Q_2\left(d\right)$ is a polynomial of one degree higher than $Q_1\left(d\right)$.
This means that, in the limit $d\to\infty$, the full decoupling is realized and only the number of degrees of freedom matters,
a situation that is reminiscent of many mean field computations at large $d$.

\subsection{Transversal vector fields on Einstein-spaces}
\label{sect:3.4}
As already discussed in section \ref{sect:3.1}, the Einstein-spaces reviewed in appendix \ref{Einstein} constitute a special class of Riemannian manifolds where the projection operators \eqref{projectors} commute with the Laplacian. In this case the basis monomials in \eqref{totalresult} are not independent but related by the geometrical identities \eqref{esc3}.
The corresponding heat-kernel coefficients are readily obtained from the general result \eqref{totalresult} by using \eqref{esc3} to express all curvature monomials in terms
of the Einstein-space basis
\be\label{S1ESS}
\begin{split}
S_{\rm 1T}\Big|_{\rm ES} = & \, \tfrac{1}{(4\pi s)^{d/2}}  \int d^dx \sqrt{g} \Big\{
(d-1) + \tfrac{(d-3)(d+2)}{6d} s \cE^1 \\
&
+ s^2 \left( \tfrac{5d^3-7d^2-58d-180}{360d^2}  \cE^2_1
+ \tfrac{d-16}{180} \cE^2_2 \right) \\
&
+ s^3 \left( \tfrac{35d^4-77d^3-604d^2-3512d-7560}{45360 d^3}  \cE^3_1
+ \tfrac{7d^2-111d-127}{7560 d}  \cE^3_2 
+ \tfrac{17d-269}{45360}  \cE^3_3
- \tfrac{d-19}{1620}  \cE^3_{4} \right)
\Big\} \, .
\end{split}
\ee
Remarkably, in this case we find heat-kernel coefficients that are \emph{finite in any dimension} $d > 0$. Evoking the Einstein-space limit the heat-kernel coefficients in \eqref{totalresult} combine in such a way that all the singularities appearing for even dimensionality $d$ cancel.

At this stage, it is illustrative to rederive \eqref{S1ESS} by using the Einstein-space condition from the very beginning. Since $[\D, \Pi_{\rm T}]|_{\rm ES} = 0$, all $S^{(n)}_{\rm 1T}$, $n \ge 1$ in the expansion \eqref{1Ttrace_expanded} vanish and
\be
S_{\rm 1T}\big|_{\rm ES} = S_{\rm 1T}^{(0)}
\ee
is exact. Exploiting that on an Einstein-space
\be\label{commfuncES}
\begin{split}
D_\a \, f\big(\Delta\big) \phi  = f\big(\Delta+\tfrac{R}{d}\big)D_\a \phi \, , \qquad
D_\a \, f\big(\Delta\big) v^\a  = f\big(\Delta-\tfrac{R}{d}\big)D_\a v^\a \, ,
\end{split}
\ee
the trace \eqref{D.1} can be cast into standard form without resorting to the expansion \eqref{S10exp}
\be\label{S1TES}
\begin{split}
S_{\rm 1T}\big|_{\rm ES} = 
 & \, \Tr_1 \Big[ \d_\m^\n \e^{-s\Delta} \Big]
+ \int_0^\infty  \Tr_1 \Big[ \e^{-t \tfrac{R}{d}} \, D_\m D^\n \e^{-(s+t)\Delta} \Big] dt \, .
\end{split}
\ee
Here the Ricci scalar in the exponential results from the summation of the multi-commutators in (\ref{S10exp}).
The curvature expansion of the traces can then, again, be found via the off-diagonal heat-kernel. The additional exponential thereby leads to an exponential
suppression of the integrand for large values of $t$, rendering the auxiliary integrals finite. This computation confirms the result \eqref{S1ESS}. Notably,
the exponential ensures that the heat-kernel coefficients appearing in $S_{\rm 1T}\big|_{\rm ES}$ are free from dimensional poles to all orders in $\cR^{(n)}$.
Moreover, the fact that the general computation restricted to the Einstein-space limit and the Einstein-space computation lead to the same finite result indicates that the origin of the dimensional poles is not due to the use of the early-time expansion of the off-diagonal heat-kernel. We will elaborate this point further in the next subsection.

We close this subsection by evaluating \eqref{S1ESS} on a spherical background. Using \eqref{spherelim} we obtain
\be\label{S1Tsphere}
\begin{split}
S_{\rm 1T}\Big|_{\rm Sphere} = \; \tfrac{1}{(4\pi s)^{d/2}}  \int d^dx \sqrt{g} \Big\{&
(d-1) + \tfrac{(d-3)(d+2)}{6d} s R
+ \tfrac{5d^4-12d^3-47d^2-186d+180}{360d^2(d-1)} s^2 R^2 \\
&
+ \tfrac{35 d^6-147d^5-331d^4-3825d^3-676d^2+10992d-7560}{45360d^3(d-1)^2} s^3 R^3
\Big\} \, , \\
\end{split}
\ee
which for $d=4$ becomes
\be
\begin{split} 
S_{\rm 1T}\Big|_{{\rm Sphere}, d=4} =  \; \tfrac{1}{(4\pi s)^2}  \int d^4x \sqrt{g} \Big\{&
3 + \tfrac{1}{4} s R - \tfrac{67}{1440} s^2 R^2 - \tfrac{4321}{362880} s^3 R^3
\Big\} \, .
\end{split}
\ee
This trace still contains the contribution of the constant scalar mode, which does not contribute to the $T$-decomposition \eqref{Tdec}.
In order to obtain the final result, this mode has to be removed by hand, leading to a modified heat-trace \cite{Lauscher:2001ya}
\be
\begin{split}
\Tr^\prime_{\rm 1T}\; \e^{-s\D}
= & \, \Tr_{\rm 1T}\; \e^{-s\D} +  \e^{s\tfrac{R}{4}} \, \\
\simeq & \, S_{\rm 1T}\Big|_{{\rm Sphere}, d=4} \; + \tfrac{1}{(4\pi)^2} \int d^4x \sqrt{g}\; \tfrac{1}{24} \, R^2\, \e^{s\tfrac{R}{4}} \, .
\end{split}
\ee
Here the second term is precisely (minus) the contribution of a constant scalar. Taking this correction term into account, the heat-kernel coefficients for transversal vectors on the sphere become
\be
c^\prime_0 = 3 \, , \quad
c^\prime_1 = \tfrac{1}{4} \, , \quad
c^\prime_2 = -\tfrac{7}{1440} \, , \quad
c^\prime_3 = -\tfrac{541}{362880} \, .
\ee
Notably, this result is in complete agreement with previous computations \cite{Codello:2008vh,Lauscher:2001ya,Machado:2007ea}.

\subsection{Regularizing the poles}

The infrared divergences of the auxiliary integrals \eqref{tInt2} appear in the heat-kernel expansion of $S_{\rm 1T}$ in the form of poles in even dimensions, as shown in table \ref{t.1}. The first divergence in $d=4$ spacetime dimensions appears at order $\cR^3$, when the boundary term $\D R$ is neglected. Notably, however, the coefficients of $\cR_9^3$ and $\cR_{10}^3$ are finite for any dimensionality.

Any attempt to regularize \eqref{totalresult} has to be implemented with care in order not to affect
the unambiguous result for Einstein spaces, since in this limit the poles cancel with a factor $(d-4)$ forming in the numerator.
If no pole is present, these contributions would otherwise go to zero.
In order to take the limit $d\rightarrow 4$, we expand in $\eps = d-4$ and make use of the formula
\be\label{dimreg}
\begin{split}
s^{n-d/2} \frac{f(d)}{d-4}
&= s^{n-2} \bigg( f^\prime(4) + f(4)\Big( \tfrac{1}{\eps}-\tfrac{1}{2}\log \big(\tfrac{s}{s_0}\big) \Big)\bigg) + \cO(\eps) \\
&= s^{n-2} \bigg( f^\prime(4) -\tfrac{1}{2} f(4) \log \big(\tfrac{s}{s_0^\prime}\big) \bigg) + \cO(\eps) \, ,
\end{split}
\ee
where the $\tfrac{1}{\eps}$ pole is absorbed into a (divergent) redefinition of the scale $s_0^\prime = s_0 \e^{2/\eps}$. In this way $s_0^\prime$ plays the role of an infrared cutoff and allows us to obtain the limit as
\be\label{totalresult4d}
\begin{split}
S_{\rm 1T}\big|_{d\rightarrow 4} 
= \frac{1}{(4\pi s)^{2}} & \int d^{4+\epsilon}x \sqrt{g} \bigg\{
3 \; + \tfrac{1}{4} \;s R\;
+ s^2 \Big( - \tfrac{1}{48} \;R^2\;
- \tfrac{7}{120} \;\cR_2^2\;
- \tfrac{1}{15} \;\cR_3^2 \Big) \\
& + s^3 \Big(
 \big( \tfrac{1363}{20160}+\tfrac{1}{30}\log\big(\tfrac{s}{s_0^\prime}\big) \big) \;\cR_1^3\;
- \big( \tfrac{67}{2016}+\tfrac{1}{60}\log\big(\tfrac{s}{s_0^\prime}\big) \big) \;\cR_2^3 \\
& \qquad
+ \big( \tfrac{41}{3456}+\tfrac{1}{144}\log\big(\tfrac{s}{s_0^\prime}\big) \big) \;\cR_3^3\;
- \big( \tfrac{263}{2880}+\tfrac{11}{360}\log\big(\tfrac{s}{s_0^\prime}\big) \big) \;\cR_4^3 \\
& \qquad
+ \big( \tfrac{233}{1512}+\tfrac{1}{45}\log\big(\tfrac{s}{s_0^\prime}\big) \big) \;\cR_5^3\;
- \big( \tfrac{397}{5040}+\tfrac{1}{90}\log\big(\tfrac{s}{s_0^\prime}\big) \big) \;\cR_6^3 \\
& \qquad
- \big( \tfrac{1}{90}-\tfrac{1}{360}\log\big(\tfrac{s}{s_0^\prime}\big) \big) \;\cR_7^3\;
- \big( \tfrac{3}{280}+\tfrac{1}{90}\log\big(\tfrac{s}{s_0^\prime}\big) \big) \;\cR_8^3 \\
& \qquad
- \tfrac{67}{15120} \;\cR_9^3\;
+ \tfrac{1}{108} \;\cR_{10}^3
\Big)
\bigg\} \, .
\end{split}
\ee
The appearance of the logarithmic terms is an immediate consequence of the dimensional regularization employed. The Seeley-deWitt expansion is still ``weakly'' valid after dimensional regularization, because the singular $\log (s)$ terms are always multiplied by  a power of $s$, and so maintain a continuous $s\rightarrow 0$ limit.\footnote{The very same happens also in the $d\to 2$ limit.}
Using this regularization scheme, in principle no information is lost by regularizing by a finite choice of $s_0^\prime$ in \eqref{totalresult4d}. In fact, it is still possible to recover the correct Einstein-space limit from \eqref{totalresult4d}, provided that the explicit dependence of $s_0^\prime$ on $\eps=d-4$ is taken into account, so that it can combine with the $d$-dependence of the Ricci- and scalar curvature tensors in their Einstein-space limit.

To demonstrate the origin of the divergence in the low momentum contribution of the inverse Laplacian in the projectors (\ref{projectors}), we consider a one-parameter family of ``projectors''
\be\label{a_parameter}
\check{\P}^{\rm T}_{\m\n} = g_{\m\n} + a D_\m \frac{1}{\D} D_\n \, ,
\ee
that are idempotent for $a=1$ only. With this definition the modified projected Laplacian can be expanded in the form
\be
\check{\Delta}_{\m\n} = (\check{\P}^{\rm T}\,\Delta\,\check{\P}^{\rm T})_{\m\n} = \D g_{\m\n} + a (2-a) D_\m D_\n + {\cal O}\!\left({\cal R}, \Delta^{-1}\right)\, ,
\ee
where all the pseudo-differential contributions due to the commutators of the inverse Laplacian with the covariant derivatives are schematically contained in ${\cal O}\!\left({\cal R}, \Delta^{-1}\right)$.
At zeroth order in the curvature, the heat-kernel of $\check{\Delta}$ is singular at $a=1$, due to the degeneracy of the operator \cite{Barvinsky:1985an,Gusynin:1997qs}.
Since for other values of $a$ $\check{\P}^{\rm T}$ is not a projector anymore, we can exclude the degeneracy, that is specifically its infinite number of zero eigenvalues corresponding to the volume of the subspace of longitudinal modes, as source of the divergence. Tracking this modification in the traces \eqref{S1Tn} reveals a simple polynomial dependence of \eqref{totalresult} on the parameter $a$. Therefore a discontinuity for the special value $a=1$ is not present. An explicit calculation in fact shows that the dimensional poles vanish only for $a=0$, reproducing the heat-kernel of the standard Laplacian.

Alternatively, we can remove the zero eigenvalue from the spectrum of the Laplacian specifically by shifting it with an infrared scale $m^2$, defining the operator
\be\label{PTmass}
\P^{\rm T,m^2}_{\m\n} = g_{\m\n}+D_\m \big(\D+m^2 \big)^{-1} D_\n \, .
\ee
The scale $m^2$ can be chosen arbitrarily small, suppressing the low momentum modes while leaving the high momentum spectrum, essentially, unaltered.
Using this modified operator in place of the projectors in (\ref{1Ttrace}), the exponentiation via (\ref{tInt0}) would be done on the regulated inverse Laplacian $(\D+m^2)^{-1}$. This accounts for an additional regularizing factor $\e^{-tm^2}$, which renders the modification of the integrals \eqref{tInt2} convergent at any order of the curvature. We conclude that the long range (infrared) modes of the pseudo-differential projection operators cause the breakdown of the heat-kernel expansion (\ref{totalresult}).
This situation is analogue to the case of a scalar field on curved spacetime, whose propagator can be defined via the heat-kernel in a local expansion only if it is massive (or otherwise IR regulated).

For practical purposes, it is conceivable to use a modified operator like \eqref{PTmass}, if $m^2$ can be identified with an infrared scale already present in a particular problem. In general however, such a procedure has the disadvantages that the limit $m^2\rightarrow 0$ is discontinuous and arbitrary powers of the dimensionless combination $s m^2$ will occur in heat-kernel traces.
Instead, we suggest to follow \eqref{dimreg} and to introduce an infrared scale in a purely dimensionally regulated setup.

\section{Summary}
\label{summary}

\begin{table}[t]
\begin{center}
\begin{tabular}{|c||c|c|c||c|}
\hline
& scalar & vector & symmetric tensor & transverse vector \\ \hline
$c_0$ & $1$               & $d_1$ & $d_2$ &  $d_{\rm 1T}$ \\[1.1ex] \hline
$c_1$ & $\frac{1}{6}$     & $\frac{d_1}{6}$ & $\frac{d_2}{6}$ & $\frac{d_{\rm 1T}}{6} - \tfrac{1}{d}$ \\[1.1ex] \hline
%
%
$c_2^1$ & $\frac{1}{72}$  & $\frac{d_1}{72}$ & $\frac{d_2}{72}$ &   $\frac{d_{\rm 1T}}{72} -\tfrac{d^2 - d + 6}{6 (d-2) d (2 + d)}$ \\[1.1ex]
$c_2^2$ & $-\frac{1}{180}$& $-\frac{d_1}{180}$ & $-\frac{d_2}{180}$ & $-\frac{d_{\rm 1T}}{180} - \tfrac{2 d^2 - 5 d  - 6 }{3 (d-2) d (2 + d)}$ \\[1.1ex]
$c_2^3$ & $\frac{1}{180}$ & $\frac{d_1}{180} - \frac{1}{12}$ & $\frac{d_2}{180} - \frac{d+2}{12}$ &  $\frac{d_{\rm 1T}}{180} - \frac{1}{12}$ \\[1.1ex] \hline
%
%
$c_3^1$ &  $\frac{1}{336}$ & $\frac{d_1}{336} + \frac{1}{120}$ & $\frac{d_2}{336} + \frac{d+2}{120}$ &  $\frac{d_{\rm 1T}}{336} + \frac{1}{120} +  \frac{8d^4-73d^3+208d^2-428d+240}{30 (d-4)(d-2)d(d+2)(d+4)}$ \\[1.1ex]
$c_3^2$ &  $\frac{1}{840}$ & $\frac{d_1}{840} - \frac{1}{30}$ & $\frac{d_2}{840} - \frac{d+2}{30}$ & $\frac{d_{\rm 1T}}{840} - \frac{1}{30} + \frac{2d^4-17d^3+42d^2+88d-320}{10 (d-4)(d-2)d(d+2)(d+4)} $ \\[1.1ex]
$c_3^3$ & $\frac{1}{1296}$ & $\frac{d_1}{1296}$ & $\frac{d_2}{1296}$ & $\frac{d_{\rm 1T}}{1296} - \frac{d^4-2d^3-4d^2+8d+288}{72(d-4)(d-2)d(d+2)(d+4)}$ \\[1.1ex]
$c_3^4$ &  $-\frac{1}{1080}$ & $-\frac{d_1}{1080}$ & $-\frac{d_2}{1080}$ & $-\frac{d_{\rm 1T}}{1080} - \frac{19d^3-82d^2+148d-1200}{180 (d-4)(d-2)d(d+4)}$\\[1.1ex]
$c_3^5$ &  $-\frac{4}{2835}$ & $-\frac{4d_1}{2835} + \frac{1}{30}$ & $- \frac{4d_2}{2835} + \frac{d+2}{30}$ & $- \frac{4d_{\rm 1T}}{2835} + \frac{1}{30} + \frac{41d^4-136d^3-44d^2-896d+960}{90 (d-4)(d-2)d(d+2)(d+4)}$ \\[1.1ex]
$c_3^6$ &  $\frac{1}{945}$ & $\frac{d_1}{945} - \frac{1}{30}$ & $\frac{d_2}{945} - \frac{d+2}{30}$ & $\frac{d_{\rm 1T}}{945} - \frac{1}{30} - \frac{29d^4-139d^3-86d^2+376d+960}{45(d-4)(d-2)d(d+2)(d+4)}$ \\[1.1ex]
$c_3^7$ &  $\frac{1}{1080}$ & $\frac{d_1}{1080} - \frac{1}{72}$ & $\frac{d_2}{1080} - \frac{d+2}{72}$ &  $\frac{d_{\rm 1T}}{1080} - \frac{1}{72}- \frac{1}{180(d-4)}$ \\[1.1ex]
$c_3^8$ &  $\frac{1}{7560}$ & $\frac{d_1}{7560} - \frac{1}{90}$ & $\frac{d_2}{7560} - \frac{d+2}{90}$ & $\frac{d_{\rm 1T}}{7560} - \frac{1}{90} + \frac{1}{45(d-4)}$ \\[1.1ex]
$c_3^9$ &  $\frac{17}{45360}$ & $\frac{17d_1}{45360} - \frac{1}{180}$ & $\frac{17d_2}{45360} - \frac{d+2}{180}$ & $\frac{17d_{\rm 1T}}{45360} - \frac{1}{180}$ \\[1.1ex]
$c_3^{10}$ & $-\frac{1}{1620}$ & $-\frac{d_1}{1620} + \frac{1}{90}$ & $-\frac{d_2}{1620} + \frac{d+2}{90}$ & $-\frac{d_{\rm 1T}}{1620} + \frac{1}{90}$ \\ \hline
\end{tabular}
\end{center}
\caption{The traced heat-kernel coefficients in the early-time expansion \eqref{earlytimeexp} on the space of scalars, vectors, symmetric $2$-tensors and transversal vectors $(1T)$, respectively. The dimensions of the field spaces are denoted by $d_1 = d , d_2 = \half d (d+1)$ and $d_{\rm 1T} = d-1$, respectively.}\label{t.1}
\end{table}

In this article, we used the deWitt-algorithm to recursively determine the off-diagonal heat-kernel expansion of a Laplace operator on a general bundle, including an arbitrary endomorphism, up to third order in the curvature tensors.  The algorithm has been implemented on computer algebra systems \cite{Mathtensor,Ricci}, making the manual handling of the huge expressions for higher orders unnecessary.
Our results generalize previous calculations \cite{Benedetti:2010nr,Decanini:2005gt,Decanini:2007gj,Anselmi:2007eq} by leaving the internal space and spacetime dimension unspecified. In particular the $\overline{A_3}$ given in eq. \eqref{A3} is, to our knowledge, the most general determination of this coefficient, since it is given without tracing over the internal space.
All the results for the traced expansion of the heat-kernel of the Laplace operator acting on scalar-, vector- and symmetric tensor fields are conveniently summarized in table \ref{t.1}.

Our results allow the systematic evaluation of the traced heat-kernel, including insertions of covariant derivative operators inside the trace.
These results become relevant for the evaluation of heat-traces of non-minimal differential operators,
for which the heat-kernel expansion is harder to obtain and known to less extent \cite{Gusev}.
In this case, an expansion and resummation of the non-minimal part allows to reduce the heat-trace to terms of the form \eqref{non-minimal-trace} \cite{Barvinsky:1985an,Saueressig:2011vn}.
This is particularly useful in the context of quantum field theory, where expressions of this form
generally
appear when a gauge-invariance is present.
As an example,
the newly computed non-minimal traces in section \ref{main:1} provide the means to extent previous approximations of the functional renormalization group flow of gauge theories and gravity in numerous ways. In particular, they will be an essential ingredient in extending the universal renormalization group machine \cite{Benedetti:2010nr} to the third order in the curvatures.

As a non-trivial application of our results, we studied the traced heat-kernel on the space of transverse vector fields satisfying $D^\m v_\m = 0$.
Notably, the Laplace operator projected on the transverse vectors subspace becomes non-local
due to the projection operators.
Owed to this non-locality, the Seeley-deWitt coefficients of the heat-trace, on a general $d$-dimensional manifold, diverge starting from order $\cR^{d/2+1}$.
The divergences are of infrared nature and appear as dimensional poles, when dimensional regularization is adopted.
Most remarkably, we showed that these singularities cancel, if the spacetime is an Einstein-space.
The reason is that the Einstein condition ensures that the standard Laplacian commutes with the projection onto the transverse subspace.

Remarkably, the occurrence of the singularities is not related to the use of the early-time expansion of the heat-kernel employed in section \ref{main:2}.
Based on power-counting arguments, we estimated that divergences are present also if a non-local expansion \eqref{bv} of the projected trace is employed.
This estimate leads to the same divergence structure in the form of dimensional poles, as that indicated by the results reported in table \ref{t.1}.
We take this as a strong indication that our poles are not an artifact of the expansion,
but rather a genuine feature of the projected traces.

We close our summary by commenting on the consequences of these divergences in the context of the functional renormalization group approach to quantum gravity.
In this context, the method of transversal field decomposition was first advocated in \cite{Lauscher:2001ya} and subsequently employed by a number of groups \cite{Codello:2008vh,Machado:2007ea,Lauscher:2001rz,Benedetti:2009rx,Benedetti:2009gn,Eichhorn:2010tb} to diagonalize the propagator.
This technique works as long as the background manifold satisfies the Einstein condition,
but our results indicate that it does not generalize to the case of generic
backgrounds.
Instead of using the transverse decomposition, one can perform a resummation of the non-minimal derivative terms appearing in the inverse propagator \cite{Barvinsky:1985an}.
This idea has already been used to reobtain the 1-loop effective action of four-derivative gravity in \cite{Saueressig:2011vn},
and can be generalized to a full non-perturbative computation of the effective action.

\section*{Acknowledgments}
%
We thank M.\ Reuter, D.\ Benedetti, A.\ Codello and R.\ Percacci for helpful discussions. The research of K.~G., F.~S., and O.~Z.\ is supported by the Deutsche Forschungsgemeinschaft (DFG)
within the Emmy-Noether program (Grant SA/1975 1-1).

\begin{appendix}

\section{Curvature monomials on Riemannian manifolds}
\label{app:1}
An essential ingredient for the early-time expansion of the
heat-kernel is a basis for the independent curvature monomials
at a given order $\cR^n$. In this appendix, we therefore
review the results of \cite{Decanini:2007gj,Fulling:1992vm,Decanini:2008pr}
which are essential for our construction. Throughout this appendix, we
will consider the case of $d$-dimensional Riemannian manifolds without boundaries.
%
\subsection{Curvature basis for general manifolds}
\label{app:1.1}
The heat-kernel expansions \eqref{earlytimeexp} and \eqref{totalresult} require a basis 
for all curvature invariants build from six or less covariant derivatives. Following
\cite{Fulling:1992vm} this basis contains 15 elements, which we choose as follows\footnote{Taking total derivatives into account, there is one more invariant at $\cO(\cR^2)$
and seven additional curvature monomials at $\cO(\cR^3)$, see eqs.\ \eqref{noboundid1} and \eqref{noboundid} below.}
\be\label{basis}
\begin{array}{lll}
\cR^0 = 1 \, , & & \\
\cR^1 = R \, , & & \\
\cR^2_1 = R^2 \, , & \cR^2_2 = R_{\m\n}R^{\m\n} \, , & \cR^2_3 =  R_{\m\n\a\b} R^{\m\n\a\b} \\
\cR^3_1 = R \square R \, , &
\cR^3_2 = R_{\m\n} \square  R^{\m\n} \, , &
\cR^3_3 = R^3 \, , \\  
\cR^3_4 = R R_{\m\n} R^{\m\n} \, , &
\cR^3_5 = R_{\m}{}^\n R_\n{}^\alpha R_\alpha{}^\m \, , &
\cR^3_6 = R_{\m\n} R_{\a\b} R^{\m\a\n\b} \, , \\
\cR^3_7 = R R_{\a\b\m\n} R^{\a\b\m\n} \, , &
\cR^3_8 = R_{\m\n} R^{\m\a\b\gamma} R^{\n}{}_{\a\b\gamma} \, , &
\cR^3_9 = R_{\m\n}{}^{\rho\sigma} R_{\rho \sigma}{}^{\a\b} R_{\a\b}{}^{\m\n} \, , \\
\cR^3_{10} = R^{\a}{}_{\m}{}^{\b}{}_{\n} R^\m{}_\rho{}^\n{}_\sigma R^{\rho}{}_\a{}^\sigma{}_\b \, .
\end{array}
\ee
Here $\square \equiv g^{\m\n} D_\m D_\n$.

Notably, the results obtained from the off-diagonal heat-kernel techniques are, a priori, not given
with respect to this basis, but also include additional curvature monomials, which are related to
the ones above through the first and second Bianchi identities
\be\label{bianchi1}
R_{\m[\n\a\b ]} = 0
\ee
and
\be\label{bianchi2}
\begin{split}
& R_{\m\n\a\b;\rho} + R_{\m\n\b\rho;\a} + R_{\m\n\rho\a;\b} = 0 \, , \quad
 R_{\m\n\a\b}{}^{;\b} = R_{\m\a;\n} - R_{\n\a;\m} \, , \quad
 R_{\m\n}{}^{;\n} = \half R_{;\m} \, .
\end{split}
\ee
A very useful collection of curvature identities implied by these identities has been given in \cite{Decanini:2008pr}, and for completeness we summarize the ones important for our construction.

At order $\cR^2$, \eqref{bianchi1} implies
\be
R^{\m\n\a\b} R_{\m\a\n\b} = \half  R_{\m\n\a\b} R^{\m\n\a\b} \, , 
\ee
while at $\cR^3$ there are three relations between different contractions of the Riemann tensor
\be
\begin{split}
R^{\m\a\n\b} R_{\m\n\r\s} R_{\a\b}{}^{\r\s} = & \, \half R_{\m\n}{}^{\rho\sigma} R_{\rho \sigma}{}^{\a\b} R_{\a\b}{}^{\m\n} \, , \\
R_{\a\b}{}^{\r\s} R^{\a\m\b\n} R_{\r\m\s\n} = & \, \tfrac{1}{4}  R_{\m\n}{}^{\rho\sigma} R_{\rho \sigma}{}^{\a\b} R_{\a\b}{}^{\m\n} \, , \\
R_{\m\a\n\b} R^{\m\r\n\s} R^\a{}_\s{}^\b{}_\r = & \,  - \tfrac{1}{4} R_{\m\n}{}^{\rho\sigma} R_{\rho \sigma}{}^{\a\b} R_{\a\b}{}^{\m\n} + R^{\a}{}_{\m}{}^{\b}{}_{\n} R^\m{}_\rho{}^\n{}_\sigma R^{\rho}{}_\a{}^\sigma{}_\b \, .
\end{split}
\ee
In addition the combination of \eqref{bianchi1} and \eqref{bianchi2} allows to derive
\be
\begin{split}
R_{\m\n\a\b} \square R^{\m\n\a\b} = & \, 4 R_{\m\n;\a\b} R^{\m\a\n\b} + 2 R_{\m\n} R^{\m}_{\a\b\gamma} R^{\n\a\b\gamma} - R_{\m\n}{}^{\rho\sigma} R_{\rho \sigma}{}^{\a\b} R_{\a\b}{}^{\m\n} \\
& \, - 4 R^{\a}{}_{\m}{}^{\b}{}_{\n} R^\m{}_\rho{}^\n{}_\sigma R^{\rho}{}_\a{}^\sigma{}_\b \, ,
\end{split}
\ee
and
\be
R_{\m\n} R^{\m\a;\n}{}_\a = \half R_{;\m\n} R^{\m\n} + R_{\m}{}^\n R_\n{}^\alpha R_\alpha{}^\m - R_{\m\n} R_{\a\b} R^{\m\a\n\b} \, .
\ee

Once the curvature monomials appear under the volume integral, the condition of working on a manifold without boundary allows us to freely integrate by parts.
At $\cR^2$ this eliminates the surface term
\be\label{noboundid1}
\int d^dx \sqrt{g} \, \square R = 0 \, ,
\ee
while at order $\cR^3$ we obtain the seven additional identities
\be\label{noboundid}
\begin{split}
\int d^dx \sqrt{g} \, \square \square R = & \, 0 \, , \\
\int d^dx \sqrt{g} \, R^{\m\n} R_{;\m\n} = & \, \half \int d^dx \sqrt{g} \, \cR^3_1 ,  \\
\int d^dx \sqrt{g} \, R_{\m\n;\a\b} R^{\m\a\n\b} = & \, \int d^dx \sqrt{g} \, \left[ \cR^3_2  - \tfrac{1}{4} \cR^3_1 - \cR^3_5 + \cR^3_6 \right] \, , \\
\int d^dx \sqrt{g} \,  R_{;\m} R^{;\m} = & \, - \int d^dx \sqrt{g} \,  \cR^3_1 \, , \\
\int d^dx \sqrt{g} \,  R_{\a\b;\m} R^{\a\b;\m} = & \, - \int d^dx \sqrt{g} \, \cR^3_2 \, , \\
\int d^dx \sqrt{g} \,  R_{\a\b;\m} R^{\a\m;\b} = & \, \int d^dx \sqrt{g} \, \left[ - \tfrac{1}{4} \cR^3_1 - \cR^3_5 + \cR_6^3 \right] \, , \\
\int d^dx \sqrt{g} \,  R_{\m\n\a\b;\r} R^{\m\n\a\b;\r} = & \, \int d^dx \sqrt{g} \, \big[ \cR^3_1 - 4 \cR^3_2 + 4 \cR^3_5 -4 \cR^3_6
- 2 \cR^3_8 + \cR^3_9 + 4 \cR^3_{10} \big] \, .
\end{split}
\ee

\subsection{Curvature basis for Einstein-spaces}
\label{Einstein}
A special class of Riemannian manifolds are Einstein-spaces, where the Ricci tensor is proportional to the metric
\be\label{esc1}
R_{\m\n} = \tfrac{1}{d} g_{\m\n} R \, .
\ee
In connection with the Bianchi-identities \eqref{bianchi2} this definition entails
\be\label{esc2}
R_{;\m} = 0 \, , \quad R_{\a\b;\m} = 0 \, , \quad R_{\a\b\g\m;}{}^{\m} = 0 \, ,
\ee
for any dimension $d \not = 2$.

As a direct consequence of these additional relations, the general basis of curvature
monomials \eqref{basis} degenerates, so that, up to $\cO(\cR^3)$, an Einstein-space has only eight
distinguished curvature monomials. Explicitly, these can be chosen to be
\be\label{Einsteinbasis}
\begin{split}
\cE^0 & = 1 \, , \\
\cE^1 & = R \, , \\
\cE^2_1 & = R^2 \, , \; \cE^2_2 = R_{\m\n\a\b} R^{\m\n\a\b} \, , \\
\cE^3_1 & = R^3 \, , \; \cE^3_2  = R R_{\m\n\a\b} R^{\m\n\a\b} \, , \;
\cE^3_3 =  R_{\m\n}{}^{\r\s} R_{\r\s}{}^{\a\b} R_{\a\b}{}^{\m\n} \, , \;
\cE^3_4 = R^{\a}{}_{\m}{}^{\b}{}_{\n} R^\m{}_\r{}^\n{}_\s R^{\r}{}_\a{}^\s{}_\b \, .
\end{split}
\ee
Using eqs.\ \eqref{esc1} and \eqref{esc2}, the additional basis elements in \eqref{basis}
can be expressed in terms of the $\cE_m^n$
\be\label{esc3}
\begin{array}{lll}
\cR^2_2 = \tfrac{1}{d} \cE^2_1 \, , & \\
\cR^3_1 =  0 \, , \quad & \cR^3_2 = 0 \, , \quad & \cR^3_4 = \tfrac{1}{d} \cE^3_1 \, , \\
\cR^3_5 = \tfrac{1}{d^2} \cE^3_1 \, , \quad & \cR^3_6 = \tfrac{1}{d^2} \cR^3_1 \, , \quad & \cR^3_8 = \tfrac{1}{d} \cE^3_2 \, .
\end{array}
\ee

Lastly, the maximally symmetric $d$-spheres $S^d$ pose a special class of Einstein-manifolds. In their case, the relations
\be\label{dsphere}
R_{\m\n} = \tfrac{1}{d} g_{\m\n} R \, , \qquad R_{\m\n\r\s} = \tfrac{1}{d(d-1)}(g_{\m\r}g_{\n\s} - g_{\m\s}g_{\n\r})R \, ,
\ee
determine all curvature invariants in terms of the constant Ricci scalar or, equivalently, the radius of the sphere. Thus, the basis of curvature monomials contains only one element at each order $\cR^n$: $R^n$. The additional basis elements contained in \eqref{Einsteinbasis}
then satisfy
\be\label{spherelim}
\begin{split}
 \cE^2_2 & \, = \tfrac{2}{d(d-1)} R^2 \, , \\
 \cE^3_2 & \, = \tfrac{2}{d(d-1)} R^3 \, , \qquad
 \cE^3_3 = \tfrac{4}{d^2(d-1)^2} R^3 \, , \qquad
 \cE^3_4 = \tfrac{d-2}{d^2(d-1)^2} R^3 \, .
\end{split}
\ee
These relations are used to compare our results to previous computations of the heat-kernel on a spherical background in subsection \ref{sect:3.4}.

\section{Auxiliary commutator identities}
\label{app:2}

This appendix gives the explicit expressions for multi-commutators that appear in the evaluation of the traces over constrained vector fields in section \ref{main:2}.
In this appendix we identify $D_\m=\nabla_\m$ and define the Laplacian $\D = - g^{\m\n} D_\m D_\n$.
Symmetrization is with unit strength $(\a\b) = \half \left( \a\b + \b\a \right)$.
In order to lighten our notation we furthermore introduce the $n$-fold commutator
\be\label{nfoldcom}
\big[ \, D_\m \, , \,  \Delta \, \big]_n \equiv \big[ \big[ \, D_\m \, , \,  \Delta \, \big]_{n-1} \, , \, \Delta] 
\ee
with $\big[ \, D_\m \, , \,  \Delta \, \big]_0 = D_\m$ and $\big[ \, D_\m \, , \,  \Delta \, \big]_1 = \big[ \, D_\m \, , \,  \Delta \, \big]$. Finally, we use $\simeq$ to indicate that the corresponding equations are exact up to terms which do not contribute to the curvature expansion in section \ref{main:2}.

\subsection{Simple commutators acting on scalar and vector fields}
The computation of the operator traces \eqref{S1Tn} requires the explicit expressions for the multi-commutators \eqref{nfoldcom} acting on scalars and vector fields up to order $n=4$. Most conveniently, the higher-order commutators can be obtained recursively from the lower order ones. For completeness, we collect the corresponding expressions in this subsection.

Up to order $n=4$, the multi-commutators acting on scalar fields are given by
\be\label{scalarcom}
\begin{split}
\big[ D_\m, \Delta \big] \; \phi = & \, R_\m{}^\a D_\a \phi \, , \\
\big[ D_\m, \D \big]_2 \phi = & \,
\left( R_\m{}^\a R_\a{}^\b - (\Delta R_\m{}^\b ) + 2 R_\m{}^{\a;\b} D_\a  \right) D_\b \, \phi \, , \\
\big[ D_\m, \D \big]_3 \phi =& \,
\Big( R_\m{}^\n R_\n{}^\a R_\a{}^\b - ( \Delta (R_\m{}^\a R_\a{}^\b) ) + 2 (R_\m{}^\a R_\a{}^\b )_{;\gamma} D^\gamma 
 \\
&
- (\Delta R_\m{}^\a) R_\a{}^\b + 4 R_{\m\a}{}^{;\b\gamma} D_\gamma D^\a\Big) D_\b \phi
+ 2 R_\m{}^{\a;\b} \big[D_\b D_\a, \Delta \big] \phi \, , \\
\big[ D_\m, \D \big]_4 \phi
\simeq & \, 4 R_\m{}^{(\a;\b)\gamma} \Bigl(
2 R^\n{}_\a D_\gamma D_\b D_\n
-2 R_\a{}^\n{}_\b{}^\lambda D_\gamma D_\lambda D_\n \\
&
+ R_\gamma{}^\sigma D_\sigma D_\a D_\b
-4 R^\sigma{}_\gamma{}^\n{}_\a D_\sigma D_\b D_\n
\Bigr)\phi + \cO(\cR^{7/2}) \, ,
\end{split}
\ee
with the commutator appearing in the third term given by 
\be
\begin{split}
\big[ D_\b D_\a, \D \big] \phi
= & \Big( 2 R^\m{}_{(\b;\a)} - R_{\a\b}{}^{;\m} + 2 R^\m{}_{(\a} D_{\b)} - 2 R_\a{}^\m{}_\b{}^\n  D_\n
\Big) D_\m \phi \, .
\end{split}
\ee
Notably, the first three commutators are exact, while in the fourth one we only displayed the terms up to $\cO(\cR^3)$.

The (contracted) commutators acting on vector fields can be calculated along the same lines. In order to 
prevent the formulas from becoming too lengthy, we thereby display terms up to $\cO(\cR^3)$ only:
\be\label{eqB6}
\begin{split}
\big[ D^\alpha, \D \big]_1 \, \phi_\alpha = & \, - D_\alpha \big( R^{\alpha \beta} \phi_\beta \big) \, , \\
\big[ D^\alpha, \D \big]_2 \, \phi_\alpha
=
& \, \Big(
2 R_{\a\b} R^{\a\m\b\n}
- R^{\m\a} R_\a{}^\n
- R^{;\m\n}
 + (\Delta R^{\m\n})
- 2 R^{\a\n;\m} D_\alpha \Big) D_\m \phi_\n \, \\
 &
 + (D_\m (R^\m{}_\a R^{\a\b})) \phi_\b
+ (D_\a \D R^{\a\b}) \phi_\b \, ,\\
\big[ D^\alpha, \D \big]_3 \, \phi_\alpha
\simeq & \, \Big( 2 R_{\a\b} R^{\a\m\b\n} - R^{\m\a} R_{\a}{}^{\n} - R^{;\m\n} + (\Delta R^{\m\n}) \Big)
\Big( R_\m{}^\r D_\r \phi_\n - 2 R^\r{}_\m{}^\s{}_\n D_\r \phi_\s \Big) \\
& \, - 2 R^{\a\n;\m} \left[D_\a D_\m, \Delta \right] \phi_\n - 4 R^{\a\n;\m\sigma} D_\sigma D_\alpha D_\m \phi_\n
\, ,\\
\big[ D^\alpha, \D \big]_4 \, \phi_\alpha \simeq & \, 
- 4 R^{\a\n;\m\b} \Big(
2 R_\m{}^\lambda D_\lambda D_\beta D_\a \phi_\n
+ R_\a{}^\lambda D_\lambda D_\beta D_\mu \phi_\n
- 4 R_\m{}^\sigma{}_\n{}^\lambda D_\beta D_\alpha D_\sigma \phi_\lambda \\
& \,
- 2 R_\a{}^\sigma{}_\n{}^\lambda D_\b D_\m D_\sigma \phi_\lambda
- 4 R_\m{}^\sigma{}_\alpha{}^\lambda D_\beta D_\sigma D_\lambda \phi_\n - 2 R_\m{}^\sigma{}_\b{}^\lambda D_\sigma D_\a D_\lambda \phi_\nu
\Big) \, .
\end{split}
\ee
These expressions are completed by the commutator appearing in the three-fold commutator
\be
\begin{split}
\big[D_\a D_\m, \D \big] \phi_\n = & \,
R_\m{}^\r D_\a D_\r \phi_\n
+ R_{\a\r} D^\r D_\m \phi_\n
- 2 R_\m{}^\r{}_\n{}^\s D_\a D_\r \phi_\s
- 2 R_\a{}^\r{}_\m{}^\s D_\r D_\s \phi_\n \\ &
- 2 R_\a{}^\r{}_\n{}^\s D_\r D_\m \phi_\s
- R_{\a\m;\r} D^\r \phi_\n
+ R_{\a\r;\m} D^\r \phi_\n
+ R_{\m\r;\a} D^\r \phi_\n \\ &
- R_{\a\n;}{}^\r D_\m \phi_\r
+ R_\a{}^\r{}_{;\n} D_\m \phi_\r
- R_{\m\n;}{}^\r D_\a \phi_\r
+ R_\m{}^\r{}_{;\n} D_\a \phi_\r \\ &
- 2 R_\m{}^\r{}_\n{}^\s{}_{;\a} D_\r \phi_\s
- R_{\m\n;}{}^\r{}_\alpha \phi_\r
+ R_\m{}^\r{}_{;\n\a} \phi_\r \, .
\end{split}
\ee
For further commutator relations, we refer to the appendix of \cite{Benedetti:2010nr}.

\subsection{Commutators involving composite operators}
\label{App.2.3}
For two continuous linear operators $X, Y$ the Baker-Campbell-Hausdorff formula allows to write
\be
\e^X \, Y \, \e^{-X} = \sum_{n=0}^\infty \frac{1}{n!} (-1)^n \big[ Y , X \big]_n \, ,
\ee
where the $n$-fold commutator has been introduced in eq.\ \eqref{nfoldcom}. Identifying $X = \mp s \D$, this formula gives an exact curvature expansion for the commutator between the exponential of a Laplacian and an arbitrary operator $Y$
\be\label{com2}
\begin{split}
\big[ \, Y \, , \, \e^{-s \D} \, \big] 
= & \, - \sum_{n=1}^\infty \frac{1}{n!} \, s^n \, \big[\, Y \, , \, \D \, \big]_n \, \e^{-s \D} \\
= & \,  \e^{-s \D} \, \left( \sum_{n=1}^\infty \frac{1}{n!} \, (-s)^n \, \big[\, Y \, , \, \D \, \big]_n  \right) \, .
\end{split}
\ee

Using the formal properties of a Laplace-transformation, \eqref{com2} can be extended to the commutator of $Y$ with a function of the Laplacian
\be\label{commfunc}
\begin{split}
\left[ Y , f(\Delta) \right] 
= & \sum_{n=1}^\infty \frac{1}{n!} (-1)^{n-1} \, \big[ \, Y \, , \, \D \, \big]_n \, f^{(n)}(\D) \\
= & \sum_{n=1}^\infty \frac{1}{n!} \, f^{(n)}(\D) \, \big[ \, Y \, , \, \D \, \big]_n \, , 
\end{split}
\ee
where $f^{(n)}$ denotes the $n$-fold derivative of $f$, and all Laplacians have been moved to the very right and very left, respectively.

Based on this result, one can obtain the curvature expansion of the projection operator \eqref{projectors} in terms of the $n$-fold commutators
\be\label{projcom}
\begin{split}
\Pi_{\rm L}{}_\m{}^\n v_\n = & - \frac{1}{\Delta} \, \sum_{n=0}^\infty \left( \frac{1}{\Delta^n} \, (-1)^n \, \left[\, D_\m \, , \, \Delta \, \right]_n \right) D^\n v_\n \\
= &
- D_\m \, \sum_{n=0}^\infty \left( \left[\, D^\n \, , \, \Delta \, \right]_n \, \frac{1}{\Delta^n} \, \right) \, \frac{1}{\Delta} \, v_\n \, .
\end{split}
\ee
Again, this expansion has the advantage that all Laplacians have been moved to the very right or very left, respectively.
The explicit expressions for the multi-commutators appearing in these expansions up to order $n=4$ are given in eqs.\ \eqref{scalarcom} and \eqref{eqB6}.
These features are essential when evaluating the operator traces including projectors in appendix \ref{App.4}.

\section{More derivatives of the off-diagonal heat-kernel}
\label{App.3}
In this appendix, we give some additional explicit formulas for the derivatives of the off-diagonal heat-kernel coefficients $A_n(x,y)$ in the coincidence limit. While interesting in their own right, these formulas allow to cross-check computer-algorithms.

The terms containing five and six symmetrized derivatives of $A_0$ read
\be\label{fivederA0}
\begin{split}
\overline{D_{(\gamma} D_\beta D_\alpha D_\nu D_{\mu)} A_0} \;=\;&
\frac{1}{3} R_{(\nu\mu;\a\b\g)} +\frac{5}{12} R_{(\gamma\beta} R_{\nu\mu;\a)} +\frac{1}{3} R_{\rho(\gamma|\theta|\beta}  R^{\rho}{}_{\nu}{}^{\theta}{}_{\mu;\a)} \, ,
\end{split}
\ee
and
\be\label{sixderA0}
\begin{split}
&\overline{D_{(\delta} D_\gamma D_\beta D_\alpha D_\nu D_{\mu)} A_0} \;=\; \\ & \qquad
\frac{5}{14} R_{(\nu\mu;\a\b\g\d)} +\frac{3}{4} R_{(\delta\gamma} R_{\nu\mu;\a\b)}
+\frac{4}{7} R_{\rho(\delta|\theta|\gamma} R^{\rho}{}_{\nu}{}^{\theta}{}_{\mu;\a\b)} 
+ \frac{15}{28} R_{\rho(\gamma|\theta|\beta;\d} R^{\rho}{}_{\nu}{}^{\theta}{}_{\mu;\a)} \\
& \qquad  +\frac{5}{8} R_{(\gamma\beta;\g} R_{\nu\mu;\a)} 
+\frac{5}{72} R_{(\delta\gamma}R_{\beta\alpha}R_{\nu\mu)} + \frac{1}{6} R_{(\delta\gamma} R_{|\rho|\beta|\theta|\alpha} R^{\rho}{}_{\nu}{}^{\theta}{}_{\mu)} \\
& \qquad
+\frac{8}{63} R^{\rho}{}_{(\delta|\theta|\gamma}R^{\theta}{}_{\beta|\lambda|\alpha}R^{\lambda}{}_{\nu|\rho|\mu)} \, .
\end{split}
\ee
The expressions for three symmetrized derivatives acting on $A_1$ and one derivative acting on $A_2$ contain both the endomorphism and curvature of the non-trivial bundle. They are given by
\be
\begin{split}
\overline{D_{(\alpha} D_\nu D_{\mu)} A_1} \;=\;&
-\frac{1}{4} E_{;(\m\n\a)}
+\frac{1}{30} R_{;(\m\n\a)}
-\frac{1}{4}E R_{(\m\n;\a)}
-\frac{1}{4} E_{;(\a} R_{\m\n)}
-\frac{3}{20} F_{\r(\m;}{}^\r{}_{\n\a)}\\ &
+\frac{1}{5} F_{\r(\a} F^\r{}_{\n;\m)}
+\frac{3}{10} F_{\r(\m;\a} F^\r{}_{\n)}
-\frac{1}{12} F_{\r(\a;}{}^\r R_{\m\n)}
-\frac{1}{30} F_{\r(\m;\a} R^\r{}_{\n)}\\ &
-\frac{1}{10} F_{\r(\a} R_{\m\n);}{}^\r
+\frac{1}{10} F_{\r(\a} R^\r{}_{\n;\m)}
-\frac{1}{15} F_{\r(\a;|\s|} R^\r{}_\m{}^\s{}_{\n)}
-\frac{1}{40} (\D R_{(\m\n})_{;\a)} \\ &
+\frac{1}{24} R R_{(\m\n;\a)}
+\frac{1}{24} R_{;(\a} R_{\m\n)}
-\frac{1}{15} R_{\r(\a} R^\r{}_{\n;\m)}
+\frac{1}{60} R_{\r\s} R^\r{}_{(\m}{}^\s{}_{\n;\a)} \\ &
+\frac{1}{60} R_{\r\s;(\a} R^\r{}_\m{}^\s{}_{\n)}
+\frac{1}{30} R^{\r\s\t}{}_{(\a} R_{|\r\s\t|\n;\m)} \, ,
\end{split}
\ee
and
\be
\begin{split}
\overline{D_\mu A_2} \;=\;&
\frac{1}{12} (\D E)_{;\m}
+\frac{1}{3} E_{;\m} E
+\frac{1}{6} E E_{;\m}
+\frac{1}{12} E_{;\r} F^\r{}_\m
+\frac{1}{12} F^\r{}_\m E_{;\r}
+\frac{1}{12} E F^\r{}_{\m;\r}\\&
+\frac{1}{12} F^\r{}_{\m;\r} E
-\frac{1}{12} E_{;\m} R
-\frac{1}{12} E R_{;\m}
+\frac{1}{60} \D (F^\r{}_{\m;\r})
-\frac{1}{60} F_{\r\m} F^{\r\s}{}_{;\s}\\&
+\frac{1}{45} F^{\r\s} F_{\r\m;\s}
+\frac{1}{30} F_{\r\s;\m} F^{\r\s}
+\frac{1}{30} F_{\r\m;\s} F^{\r\s}
+\frac{1}{45} F^{\r\s} F_{\r\s;\m}
-\frac{1}{60} F^\r{}_{\s;\r} F^\s{}_\m \\&
-\frac{1}{36} F^\r{}_{\m;\r} R
-\frac{1}{30} F^\r{}_\m R_{;\r}
+\frac{1}{30} F^{\r\s} R_{\r\m;\s}
-\frac{1}{90} F_{\r\m;\s} R^{\r\s}
+\frac{1}{180} F^\r{}_{\s;\r} R^\s{}_\m\\&
-\frac{1}{45} F_{\r\s;\g} R_\m{}^{\r\s\g}
-\frac{1}{60} (\D R)_{;a} + \frac{1}{72} R R_{;\m}
-\frac{1}{180} R^{\r\s}R_{\r\s;\m}
+\frac{1}{180} R^{\r\s\tau\kappa} R_{\r\s\tau\kappa;\m} \, ,
\end{split}
\ee
respectively. Recall that $E$ and $F_{\m\n}$ may be matrix-valued with respect to the internal bundle and therefore, in general, do not commute.

The coefficients $\overline{D_{(\m} D_\n D_\a D_{\b)} A_1}$ and $\overline{D_{(\a} D_{\b)} A_2}$ which, following table \ref{Table.1} also enter into the recursive construction of $\overline{A_3(x)}$ are not given, since they are very lengthy expressions and therefore of little practical value when written explicitly. If needed these coefficients may be computed from a supplementary Mathematica file, which will be provided on request.

\section{Traces containing projection operators: intermediate results}
\label{App.4}
In this appendix, we collect the intermediate results entering the computation of the traced heat-kernel on the space of transversal vector fields in section \ref{main:2}. In this course, we will give the explicit expressions for the  $S^{(n)}_{\rm 1T}$ defined in \eqref{S1Tn} in the following subsections.
\subsection{Evaluating $S^{(0)}_{\rm 1T}$}
\label{App.4.1}
Substituting the explicit expression for $\Pi_{\rm T}$ given in \eqref{projectors} into $S^{(0)}_{\rm 1T}$ and employing the Schwinger-trick the trace can be written as
\be\label{D.1}
S_{\rm 1T}^{(0)} = \Tr_1 \Big[ \d_\m^\n \e^{-s\Delta} \Big] + \int_0^\infty \Tr_1 \Big[ D_\m \e^{-t\Delta} D^\n \e^{-s\Delta} \Big] dt \, .
\ee
Here the first trace is a standard heat-trace on the space of unconstrained vector fields and is readily evaluated by substituting the heat-kernel coefficients given in the second column of table \ref{t.1} into \eqref{earlytimeexp}.

The second trace is evaluated via the off-diagonal heat-kernel. We first combine the two exponentials using the Baker-Campbell-Hausdorff
formula
\be\label{S10exp}
\Tr_1 \Big[ D_\m \e^{-t\Delta} D^\n \e^{-s\Delta} \Big] = \sum_{n=0}^\infty \, \frac{1}{n!} \, (-t)^n \, T^{(n)} \, ,
\ee
where
\be\label{S10expcoeff}
\begin{split}
T^{(n)} = & \, \Tr_1 \Big[ [D_\m, \D]_n \, D^\n \e^{-(s+t)\Delta} \Big] \, , \\
\end{split}
\ee
and $[D_\m, \D]_n$ denotes the $n$-fold commutator \eqref{nfoldcom}. The commutators appearing in \eqref{S10expcoeff} ensure that \eqref{S10exp} constitutes a curvature expansion of the trace. 
Following the argument at the beginning of subsection \ref{main:2.3}, all terms contributing to the basis \eqref{basis} are generated by the first five terms of this expansion, so that we truncate \eqref{S10exp} at $n=4$. Substituting the commutators \eqref{scalarcom}, it is a straightforward application of the off-diagonal heat-kernel to evaluate the $T^{(n)}$. We find
\be
\begin{split}
T^{(0)} = & \frac{1}{(4\pi(s+t))^{d/2}} \int d^dx \sqrt{g} \, \bigg[ \tfrac{-d}{2(s+t)} - \tfrac{4+d}{12} \cR^1
+ (s+t) \Big(-\tfrac{d+8}{144}\cR^2_1 + \tfrac{d-34}{360}\cR_2^2 - \tfrac{d-4}{360}\cR^2_3 \Big) \\
& + \frac{(s+t)^2}{24} \Big(-\tfrac{5d+12}{140}\cR_1^3 - \tfrac{d+36}{70}\cR_2^3 - \tfrac{d+12}{108}\cR_3^3 + \tfrac{d-30}{90}\cR_4^3 + \tfrac{16d+345}{945}\cR_5^3 \\
& \qquad\qquad - \tfrac{4d+207}{315}\cR_6^3 - \tfrac{d}{90}\cR_7^3 - \tfrac{d-174}{630}\cR_8^3 - \tfrac{17d-102}{3780}\cR_9^3 + \tfrac{d-6}{135}\cR_{10}^{3} \Big)
\bigg] \, , \\
T^{(1)} = & -  \frac{1}{(4\pi(s+t))^{d/2}} \int d^dx \sqrt{g} \, \bigg[\tfrac{1}{2(s+t)}\cR^1 + \tfrac{1}{12}\cR^2_1 + \tfrac{1}{3}\cR_2^2 \\
& + \frac{s+t}{24} \Big(-\tfrac{6}{5}\cR^3_1 + \tfrac{8}{5}\cR^3_2 + \tfrac{1}{6}\cR_3^3 + \tfrac{19}{15}\cR^3_4 - \tfrac{22}{15}\cR^3_5 + \tfrac{56}{15}\cR^3_6 + \tfrac{1}{15}\cR^3_7 - \tfrac{4}{15}\cR^3_8 \Big)
\bigg]\, ,  \\
T^{(2)} = & \frac{1}{(4\pi(s+t))^{d/2}} \int d^dx \sqrt{g} \, \bigg[ -\tfrac{1}{2(s+t)}\cR_2^2 +\frac{1}{12} \Big( 3\cR^3_1 - \cR^3_4 + 2\cR^3_5 - 6\cR^3_6 \Big)
\bigg] \, , \\
T^{(3)} = & \frac{1}{(4\pi)^{d/2} (s+t)^{d/2+1}} \int d^dx \sqrt{g} \, \bigg[ \half \cR^3_1 +\half \cR^3_2 + \half \cR^3_5 - \cR^3_6 \bigg] \, , \\
T^{(4)} = & \frac{1}{(4 \pi)^{d/2} (s+t)^{d/2+2}} \int d^dx \sqrt{g} \, \bigg[ \half \cR^3_1 + \cR^3_2 + \cR^3_5 - \cR^3_6 \bigg] \, . \,
\end{split}
\ee
Based on these results, the auxiliary $t$-integration in \eqref{D.1} can be executed. Adding the vector-trace contribution, we arrive at the final result of the form
\be\label{S1T0final}
S_{\rm 1T}^{(0)} = \frac{1}{(4\pi s)^{d/2}} \int d^dx \sqrt{g} \left[ c^0 \cR^0 + s \, c^1 \, \cR^1 + s^2 \sum_{i=1}^3 c_i^2 \, \cR_i^2 + s^3 \sum_{i=1}^{10} c_i^3 \, \cR^3_i\right] \, ,
\ee
with the $d$-dependent coefficients given in the second column of table \ref{t.2}.
\begin{table}[t]
\begin{center}
\begin{tabular}{|c||c|c|c|}
\hline
& $S_{\rm 1T}^{(0)}$ & $S_{\rm 1T}^{(1)}$ &  $S_{\rm 1T}^{(2)}$ \\ \hline
$c^0$ &  $d-1$ & $0$ & $0$  \\[1.1ex] \hline
$c^1$ &  $\frac{d}{6}-\tfrac{d+6}{6d} $ & $0$ & $0$ \\[1.1ex] \hline
$c^2_1$  & $\frac{d}{72}- \tfrac{1}{72} \tfrac{d+10}{d-2} $ & $-\tfrac{1}{d(d+2)}$ & $-\tfrac{1}{d(d+2)}$  \\[1.1ex]
$c_2^2$  & $-\frac{d}{180}+ \tfrac{d^2-32d+180}{180 d (d-2)} $ & $\tfrac{1}{(d+2)}$ & $\tfrac{1}{(d+2)}$ \\[1.1ex]
$c^2_3$  & $\frac{d-1}{180} - \frac{1}{12}$ & $0$ &  $0$  \\[1.1ex] \hline
$c^3_1$  & $\frac{d}{336} + \frac{1}{120} -\tfrac{5 d^2+ 32 d+ 464}{1680 ( d-4) (d+2)}$ & $ - \tfrac{d^3+4d^2+24d-24}{6(d-2)d(d+2)(d+4)}$ & $\tfrac{d^2- 14 d+8}{2 (d-2) d (d+2) (d+4)}$ \\[1.1ex]
$c^3_2$  & $\frac{d}{840} - \frac{1}{30} -\tfrac{d^3+ 40 d^2- 64 d-1120}{840 (d-4) d (d+2)}$ & $  \tfrac{d^2+4d-24}{6(d-2)d(d+4)}$ & $-\tfrac{4}{(d-2) (d+2) (d+4)}$ \\[1.1ex]
$c^3_3$  & $\frac{d}{1296} -\tfrac{d+14}{1296 (d-4)}$ & $ - \tfrac{1}{6(d-2)d}$ & $-\tfrac{d^2+12d-4}{6 ( d-2) d ( d+2) ( d+4)}$ \\[1.1ex]
$c^3_4$  & $-\frac{d}{1080} + \tfrac{d^2- 30 d+236}{1080 (d-4 ) (d-2)}$ & $\tfrac{d^2+d+10}{6(d-2)d(d+2)}$ & $\tfrac{d^3+14d^2-4d+40}{6(d-2)d(d+2)(d+4)}$ \\[1.1ex]
$c^3_5$  & $-\frac{4d}{2835} + \frac{1}{30}+ \tfrac{ 16 d^4 + 377 d^3 - 1954 d^2+ 2272 d-30240}{11340 (d-4) (d-2) d (d+2)}$ & $- \tfrac{d^2+8d+32}{6d(d+2)(d+4)}$ & $-\tfrac{d^2-2d+16}{3d(d+2)(d+4)}$  \\[1.1ex]
$c^3_6$  & $\frac{d}{945} - \frac{1}{30}-\tfrac{4 d^2+ 223 d-1460}{3780 (d-4) (d+2)}$ & $ \tfrac{d^3-12d-32}{3(d-2)d(d+2)(d+4)}$ & $-\tfrac{4 (d^2 + 8 )}{3 (d-2) d (d+2) (d+4)}$  \\[1.1ex]
$c^3_7$ & $\frac{d}{1080} - \frac{1}{72}-\tfrac{d+2}{1080 (d-4)}$ & $0$ & $0$  \\[1.1ex]
$c^3_8$  & $\frac{d}{7560} - \frac{1}{90} -\tfrac{d-172}{7560 (d-4)} $ & $0$ & $0$  \\[1.1ex]
$c^3_9$  & $\frac{17(d-1)}{45360} - \frac{1}{180}$ & $0$ & $0$  \\[1.1ex]
$c^3_{10}$ & $-\frac{d-1}{1620} + \frac{1}{90}$ & $0$ & $0$  \\ \hline
\end{tabular}
\end{center}
\caption{Traced heat-kernel coefficients of the partial traces $S_{\rm 1T}^{(n)}$ entering the computation in section \ref{main:2.3}.}\label{t.2}
\end{table}

%
\subsection{Evaluating $S^{(1)}_{\rm 1T}$}
\label{App.4.2}
We proceed with the evaluation of $S^{(1)}_{\rm 1T}$ defined in \eqref{S1Tn}. Substituting the commutator \eqref{firstcom} and using the orthogonality of the projectors $\Pi_{\rm T} \cdot \Pi_{\rm L} = 0$, this trace becomes
\be\label{S1int}
S^{(1)}_{\rm 1T} = - \Tr_1 \left[ \Pi_{\rm T}{}_\m{}^\a \, R_\a{}^\b \, \Pi_{\rm L}{}_\b{}^\n \e^{-s \D} \right] \, .
\ee
In order to cast this expression into standard form, we express the projection operators via \eqref{projcom}, where the inverse Laplacians already appear to the very left and very right of the commutator insertions. Since \eqref{S1int} already contains one explicit power of the curvature, the series of higher-order commutator can be terminated at order $n=3$:
\be
S^{(1)}_{\rm 1T} \simeq  \Tr_1 \left[ \left(\delta_\m{}^\a  + \frac{1}{\Delta} \, \sum_{n=0}^3 \left( \, (-\Delta)^{-n} \, \left[\, D_\m \, , \, \Delta \, \right]_n \right) D^\a \right) \, R_\a{}^\b \, D_\b \left(  \, \sum_{n=0}^3 \left[\, D^\n \, , \, \Delta \, \right]_n \, \frac{1}{\Delta^n} \, \right) \, \frac{1}{\Delta} \e^{-s \D} \right] \, .
\ee
Collecting the inverse powers of $\D$ and employing the Schwinger-trick, $S^{(1)}_{\rm 1T}$ assumes the form \eqref{tInt1}. Substituting the explicit expressions for the commutators \eqref{scalarcom} and \eqref{eqB6}, the trace becomes accessible by the off-diagonal heat-kernel.
Evaluating all subtraces with Mathematica, the final result for $S_{\rm 1T}^{(1)}$ is of the form
\be\label{S1T1final}
S_{\rm 1T}^{(1)} = \frac{s}{(4\pi s)^{d/2}} \int d^dx \sqrt{g} \left[  \sum_{i=1}^3 c_i^2 \, \cR_i^2 + s \sum_{i=1}^{10} c_i^3 \, \cR^3_i\right] \, ,
\ee
with the expansion coefficients given in the third column of table \ref{t.2}. One can verify that $S_{\rm 1T}^{(1)}$ vanishes on an Einstein-space, since $[\Delta,\Pi_{\rm T}] = 0$ for this special case. We find that the coefficients $c^3_n$ remain finite in $d=4$, so that they do not contribute to the leading divergences present in $S_{\rm 1T}^{(0)}$.

\subsection{Evaluating $S^{(2)}_{\rm 1T}$}
\label{App.4.3}
The $S^{(2)}_{\rm 1T}$-trace \eqref{S1Tn} naturally decomposes into two subtraces containing the product of two single commutators \eqref{firstcom} and the double-commutator $[\D,[\D, \Pi_{\rm T}]]$. The latter can conveniently be constructed recursively from \eqref{firstcom}. The operator insertions in these traces are then of the schematic form $\D^{-n} \cO \D^{-n^\prime} \cO^\prime \D^{-n^{\prime\prime}}$. In order to collect all Laplace operators in a single function one has to take into account that the curvature tensors are not covariantly constant and thus do not commute with the (inverse) Laplacians. These terms can be taken into account via the commutation relation
\be\label{DRiccom}
\Big[ \D^{-1} , R_{\m\n} \Big] v^\n \simeq   \left( 2 R_{\m\n;\a} D^\a -(\D R_{\m\n}) + 4 R_{\m\n;\a\b} D^\b D^\a \D^{-1} \right) \D^{-2} v^\n \, + \cO(D^3\cR) \, ,
\ee
which captures all terms contributing to the required order.

With this formula, the explicit results are found via Mathematica and read
\be\label{S1T2a}
\begin{split}
& {\rm Tr}_1 \, \Pi_{\rm T} \, \left[ \, \Delta \, , \, \Pi_{\rm T} \, \right]^2  \, \e^{-s  \Delta} = \\
& \qquad  \tfrac{1}{(4\pi s)^{d/2}} \,  \int d^dx \sqrt{g}  \, \bigg[
\tfrac{1}{d(d+2)} \left( \cR^2_1 - d \cR^2_2 \right) \\ & \qquad \quad
+ \tfrac{s}{(d-2)d(d+2)} \Big[ \tfrac{d^3+d^2+6d-8}{2(d+4)} \cR^3_1 - \tfrac{2(d^2-8)}{d+4} \cR^3_2 + \tfrac{d^2+20}{6(d+4)} \cR^3_3 \\ & \qquad \qquad
- \tfrac{d^3-4d^2+32d+40}{6(d+4)} \cR^3_4 + \tfrac{d^3+8d^2-4d-32}{3(d+4)} \cR^3_5 - \tfrac{3d^3+2d^2-24d-32}{3(d+4)} \cR^3_6
\Big] \bigg] \, ,
\end{split}
\ee
and
\be\label{S1T2b}
\begin{split}
& {\rm Tr}_1 \, \Pi_{\rm T} \, \left[ \, \Delta \, , \, \left[ \, \Delta \, , \, \Pi_{\rm T} \, \right] \right]  \, \e^{-s  \Delta} = \\
& \qquad  \tfrac{1}{(4\pi s)^{d/2}} \,  \int d^dx \sqrt{g}  \, \bigg[ - \tfrac{2}{d(d+2)} \left( \cR^2_1 - d \cR_2^2 \right) \\ & \qquad \qquad
- \tfrac{s}{6(d-2)d(d+2)} \Big[
\tfrac{3(d^3+20d-16)}{d+4} \cR^3_1 -\tfrac{12(d-4)(d+2)}{d+4} \cR^3_2 + 2(d+2) \cR_3^3 \\ & \qquad \qquad \qquad
-2(d^2+d+10) \cR^3_4 + \tfrac{4(d-2)(d^2 +4d+16)}{d+4} \cR^3_5 - \tfrac{2(d-4)(3d^2+10d+16)}{d+4} \cR^3_6
\Big]
\bigg] \,  ,
\end{split}
\ee
respectively.

The final result for $S^{(2)}_{\rm 1T}$ is the sum of \eqref{S1T2a} and \eqref{S1T2b} and has the curvature expansion
\be\label{S1T2final}
S_{\rm 1T}^{(2)} = \frac{1}{(4\pi s)^{d/2}} \int d^dx \sqrt{g} \left[  \sum_{i=1}^3 c_i^2 \, \cR_i^2 + s \sum_{i=1}^{10} c_i^3 \, \cR^3_i\right] \, ,
\ee
with the coefficients given in the fourth column of table \ref{t.2}.

\subsection{Evaluating $S^{(3)}_{\rm 1T}$ and $S^{(4)}_{\rm 1T}$}
\label{App.4.4}
The contributions of $S^{(3)}_{\rm 1T}$ and $S^{(4)}_{\rm 1T}$ can be computed along the lines of the previous subsections. In order to present the final result, it is convenient to introduce the abbreviations
\be
\begin{split}
\cC_1 \equiv \frac{1}{(4\pi s)^{d/2}} \, \frac{1}{d(d+2)(d+4)} \, \int d^dx \sqrt{g}  \, & \Big[
d(d-1) \cR^3_1 + (d^2-8) \cR_2^3 - 2 \cR^3_3 \\
&  + 3d \cR^3_4 + d(d+4) \cR^3_5 - 2d(d+2) \cR^3_6
\Big] \, ,
\end{split}
\ee
and
\be
\cC_2 \equiv \frac{1}{(4\pi s)^{d/2}} \frac{1}{(d+2)(d+4)} \int d^dx \sqrt{g} \Big[
(d-6) \cR^3_1 + 2 (d+2) \cR^3_2 + 2 d \cR^3_5 - 2d \cR^3_6
\Big] \, .
\ee
Notably both of these combinations vanish identically on an Einstein-space. The subtraces appearing in $S_{\rm 1T}^{(3)}$ are given by
\be
\begin{split}
2\,{\rm Tr}_1  \Big[ \,\Pi_{\rm T} \, \left[ \, \Delta \, , \, \Pi_{\rm T} \, \right] \, \left[ \, \Delta \, , \, \left[ \, \Delta \, , \, \Pi_{\rm T} \, \right] \right] \, \e^{-s  \Delta} \Big] = & \,2\, \cC_1 \, , \\
{\rm Tr}_1  \Big[ \,\Pi_{\rm T} \left[ \, \Delta \, , \, \left[ \, \Delta \, , \, \Pi_{\rm T} \, \right] \right] \left[ \, \Delta \, , \, \Pi_{\rm T} \, \right] \, \e^{-s  \Delta} \Big] = & \, - \cC_1 \, , \\
{\rm Tr}_1 \Big[  \,\Pi_{\rm T} \left[ \, \Delta \, , \,\left[ \, \Delta \, , \, \left[ \, \Delta \, , \, \Pi_{\rm T} \, \right] \right] \right] \, \e^{-s  \Delta} \Big] = & \, - \cC_2 \, , \\
{\rm Tr}_1 \Big[ \, \Pi_{\rm T}  \left[ \, \Delta \, , \, \Pi_{\rm T} \, \right]^3 \, \e^{-s  \Delta} \Big] =  & \, 0 \, .
\end{split}
\ee
Substituting these results into \eqref{S1Tn} yields
\be\label{S1T3final}
S_{\rm 1T}^{(3)} = \cC_1 - \cC_2 \, .
\ee
For the traces contributing to $S_{\rm 1T}^{(4)}$ we find
\be
\begin{split}
3\,{\rm Tr}_1 \Big[ \, \Pi_{\rm T} \, \left[ \, \Delta \, , \, \Pi_{\rm T} \, \right] \left[ \, \Delta \, , \, \left[ \, \Delta \, , \, \left[ \, \Delta \, , \, \Pi_{\rm T} \, \right] \right] \right] \, \e^{-s  \Delta} \Big] = &\, 3 \, s^{-1} \, \cC_2 \, , \\
3\,{\rm Tr}_1 \Big[ \, \Pi_{\rm T} \,   \left[ \, \Delta \, , \, \left[ \, \Delta \, , \, \Pi_{\rm T} \, \right] \right]^2  \, \e^{-s  \Delta} \Big]
 = & \,-3 \, s^{-1} \, \cC_2 \, , \\
{\rm Tr}_1 \Big[ \, \Pi_{\rm T} \, \left[ \, \Delta \, , \, \left[ \, \Delta \, , \, \left[ \, \Delta \, , \, \Pi_{\rm T} \, \right] \right] \right] \left[ \, \Delta \, , \, \Pi_{\rm T} \, \right] \, \e^{-s  \Delta} \Big] = & \, s^{-1} \, \cC_2 \, , \\
{\rm Tr}_1 \Big[ \, \Pi_{\rm T} \, \left[ \Delta \, , \, \left[ \, \Delta \, , \, \left[ \, \Delta \, , \, \left[ \, \Delta \, , \, \Pi_{\rm T} \, \right] \right] \right] \right] \, \e^{-s  \Delta} \Big]
= & \, 2 \, s^{-1} \, \cC_2 \, .
\end{split}
\ee
The final result for $S_{\rm 1T}^{(4)}$ is the sum of these subtraces and reads
\be\label{S1T4final}
S_{\rm 1T}^{(4)} = 3 \, s^{-1} \, \cC_2 \, .
\ee
%
\end{appendix}



\end{document}